\documentclass[namedreferences]{solarphysics}

\usepackage[figuresright]{rotating}
\usepackage[hyperref,optionalrh,solaromanenum]{spr-sola-addons}
\usepackage{graphicx}
\usepackage{txfonts}
\usepackage{natbib}
\usepackage{amssymb}
\usepackage{color}
\usepackage{booktabs}

\usepackage{caption}
\DeclareCaptionLabelSeparator{myspace}{\rule{4pt}{0pt}}
\hypersetup{
    final=true,
    pageanchor=true,
    colorlinks=true,
    breaklinks=true,
    linkcolor=blue,
    citecolor=blue,
    urlcolor=blue,
    pdfpagemode=UseNone,
    pdftitle={Image Quality in High-resolution and High-cadence Solar Imaging},
    pdfauthor={Denker et al.},
    pdfsubject={solar physics},
    pdfkeywords={granulation, sunspots, instrumental effects,
        instrumentation and data management}}

\newcommand\phm{\phantom{$-$}}
\newcommand\phn{\phantom{0}}

\newcommand\arcsec{\ensuremath{^{\prime\prime}}}

\begin{document}

\begin{article}

%
%

\begin{opening}

%
%

\title{Image Quality in High-resolution and High-cadence Solar Imaging}

\author[addressref={aff1}, corref, email={cdenker@aip.de}]
    {\inits{C.\ }\fnm{C.\ }\lnm{Denker}\orcid{0000-0002-7729-6415}}
\author[addressref={aff1,aff2}]
    {\inits{E.\ }\fnm{E.\ }\lnm{Dineva}\orcid{0000-0002-4645-4492}}
\author[addressref={aff1}]
    {\inits{H.\ }\fnm{H.\ }\lnm{Balthasar}\orcid{0000-0002-4739-1710}}
\author[addressref={aff1}]
    {\inits{M.\ }\fnm{\\M.\ }\lnm{Verma}\orcid{0000-0003-1054-766X}}
\author[addressref={aff1}]
    {\inits{C.\ }\fnm{C.\ }\lnm{Kuckein}\orcid{0000-0002-3242-1497}}
\author[addressref={aff1,aff2}]
    {\inits{A.\ }\fnm{A.\ }\lnm{Diercke}\orcid{0000-0002-9858-0490}}
\author[addressref={aff3,aff1,aff2}]
    {\inits{S.J.\ }\fnm{\\S.J.\ }\lnm{Gonz\'alez
    Manrique}\orcid{0000-0002-6546-5955}}

\address[id=aff1]{Leibniz-Institut f{\"u}r Astrophysik Potsdam (AIP), 
    An der Sternwarte 16, 14482 Potsdam, Germany}
\address[id=aff2]{Universit{\"a}t Potsdam,
    Institut f{\"u}r Physik und
    Astronomie, Karl-Liebknecht-Stra{\ss}e 24/25,
    14476~Potsdam, Germany}
\address[id=aff3]{Astronomical Institute of the Slovak Academy of Sciences, 
    05960, Tatransk{\'a} Lomnica, Slovak Republic}

\runningauthor{C.\ Denker \textit{et al.}}
\runningtitle{High-resolution and High-cadence Solar Imaging}

%
%

\begin{abstract}
Broad-band imaging and even imaging with a moderate bandpass (about 1~nm) 
provides a ``photon-rich'' environment, where frame selection (``lucky 
imaging'') becomes a helpful tool in image restoration allowing us to perform a 
cost-benefit analysis on how to design observing sequences for high-spatial 
resolution imaging in combination with real-time correction provided by an 
adaptive optics (AO) system. This study presents high-cadence (160~Hz) G-band 
and blue continuum image sequences obtained with the 
\textit{High-res\-o\-lu\-tion Fast Imager} (HiFI) at the 1.5-meter 
\textit{GREGOR} solar telescope, where the speckle masking technique is used to 
restore images with nearly diffraction-limited resolution. HiFI employs two 
synchronized large-format and high-cadence sCMOS detectors. The Median Filter 
Gradient Similarity (MFGS) image quality metric is applied, among others, to 
AO-corrected image sequences of a pore and a small sunspot observed on 2017 
June~4 and~5. A small region-of-interest, which was selected for fast imaging 
performance, covered these contrast-rich features and their neighborhood, which 
were part of active region NOAA~12661. Modifications of the MFGS algorithm 
uncover the field- and structure-dependency of this image quality metric. 
However, MFGS still remains a good choice for determining image quality without 
\textit{a priori} knowledge, which is an important characteristic when 
classifying the huge number of high-resolution images contained in data 
archives. In addition, this investigation demonstrates that a fast cadence and 
millisecond exposure times are still insufficient to reach the coherence time of 
daytime seeing. Nonetheless, the analysis shows that data acquisition rates 
exceeding 50~Hz are required to capture a substantial fraction of the best 
seeing moments, significantly boosting the performance of \textit{post-facto} 
image restoration.
\end{abstract}
\end{opening}

%
%

\section{Introduction}\label{SEC1} 

Overcoming the deleterious effects of Earth's turbulent atmosphere still poses a 
challenge in ground-based nighttime astronomy and solar observations. The 
problem of obtaining high-resolution images has been tackled using different 
approaches, \textit{e.g.} extended surveys to select the best observatory sites, 
real-time wavefront correction with AO, frame selection or lucky imaging, and 
advanced image restoration techniques.

The 1.5-meter \textit{GREGOR} telescope \citep{vonderLuehe2001, Volkmer2010b, 
Denker2012, Kneer2012, Schmidt2012} is located at Observatorio del Teide, 
Iza\~na, Tenerife, Spain. In the late 1960s, the Joint Organization for Solar 
Observations (JOSO) began a solar site survey, which initially included 40 
candidate sites. In 1979, a more thorough survey was carried out on La Palma and 
Tenerife identifying mountain locations with excellent seeing characteristics 
\citep{Brandt1982}. The evaluation of both sites was based on rms-contrast of 
granulation, image sharpness, image motion, power spectra to derive the 
Fried-parameter $r_0$ \citep{Fried1965, Fried1974}, and clear daytime fraction 
per year (about 3000 hours). The results of the JOSO site survey eventually led 
to the construction of the German solar telescopes on Tenerife 
\citep{Schroeter1985, vonderLuehe1998}. Recently, \citet{Sprung2016} presented a 
comprehensive study of the temporal evolution and the local seeing conditions at 
the \textit{GREGOR} site.

Correlation trackers \citep{vonderLuehe1989, Schmidt1995, Ballesteros1996} were 
the first attempts to remove rigid image motion and to stabilize time-series 
data in real-time. They were soon surpassed by AO systems \citep{Acton1992, 
Rimmele2000} providing instantaneous high-order wavefront correction. Today, all 
major ground-based solar telescopes are equipped with AO systems, \textit{e.g.} 
the \textit{Dunn Solar Telescope} in New Mexico \citep{Rimmele2003}, the 
\textit{Swedish Solar Telescope} at La Palma \citep{Scharmer2003}, the 
\textit{Goode Solar Telescope} at Big Bear Solar Observatory 
\citep[\textit{e.g.}][]{Rimmele2004, Denker2007b}, the \textit{New Vacuum Solar 
Telescope} at Fuxian Solar Observatory \citep{Rao2016}, and at Tenerife the 
\textit{GREGOR} solar telescope \citep{Berkefeld2012} and the \textit{Vacuum 
Tower Telescope} \citep{vonderLuehe2003}. Even for the balloon-borne 
\textit{Sunrise} telescope, well above the seeing-prone atmosphere, a low-order 
AO system provided image stabilization and wavefront correction 
\citep{Berkefeld2010}.

In parallel to the invention of correlation trackers, frame selection was 
introduced to solar observations \citep{Scharmer1989, Scharmer1991, Kitai1997}, 
which exploits the fact that for short exposure times (a few milliseconds) the 
wavefront aberrations are constant. Therefore, ``freezing'' the seeing and 
picking the best images in a time interval, which is short compared to the 
evolution time-scale of solar features (a few seconds to a minute at most), 
yields high-quality time-series. The first implementations of frame selection 
systems were based on 8-bit video technology and frame grabbers achieving data 
acquisition rates of 20\,--\,60~Hz. Grabbing images was triggered either by 
external seeing monitors or by using the high-pass-filtered analog video signal 
itself. The next improvements concerned synchronized cameras and saving larger 
sets of images for \textit{post-facto} image restoration. In addition, 
minimizing the gap between successive exposures (\textit{i.e.} minimizing the 
seeing-induced polarimetric cross-talk) and the possibility to store larger 
image sets led to digital magnetograph systems capable of obtaining magnetograms 
with sub-arcsecond resolution \citep{Lundstedt1991, Wang1998b}. The transition 
to CCD camera systems with large digitization depths and improved noise 
characteristics was slow and had the largest impact in nighttime astronomy, 
where lucky imaging and simple shift-and-add techniques \citep{Law2006, Law2009, 
Mackay2013} significantly enhanced the imaging power of large-aperture 
telescopes, especially in the visible wavelength regime. Nowadays, the progress 
in large-format and high-cadence sCMOS detectors \citep[\textit{e.g.}][for lucky 
imaging]{Qiu2013, Steele2016} offers relatively low-cost detector systems 
combining the best traits of video and CCD camera systems.

Consequently, the availability of new high-cadence imaging systems and advances 
in quantifying image quality and in turn seeing conditions, \textit{e.g.} with 
the MFGS method \citep{Deng2015}, motivated this study. The goals are i) to 
compare different MFGS implementations, ii) to evaluate potential performance 
gains in image restoration, when increasing the image acquisition rate, iii) to 
perform a cost-benefit analysis on how to design observing sequences for 
high-spatial resolution imaging, iv) to explore the corresponding parameter 
space and raise awareness for the impacts (benefits and drawbacks) of frame 
selection in image restoration, and v) to assess ramifications for the next 
generation of large-aperture solar telescopes and for archives of 
high-resolution, ground-based solar imaging data.

%
%

\section{Observations}\label{SEC2} 

\begin{figure}[t]
\centering
\includegraphics[width=\textwidth]{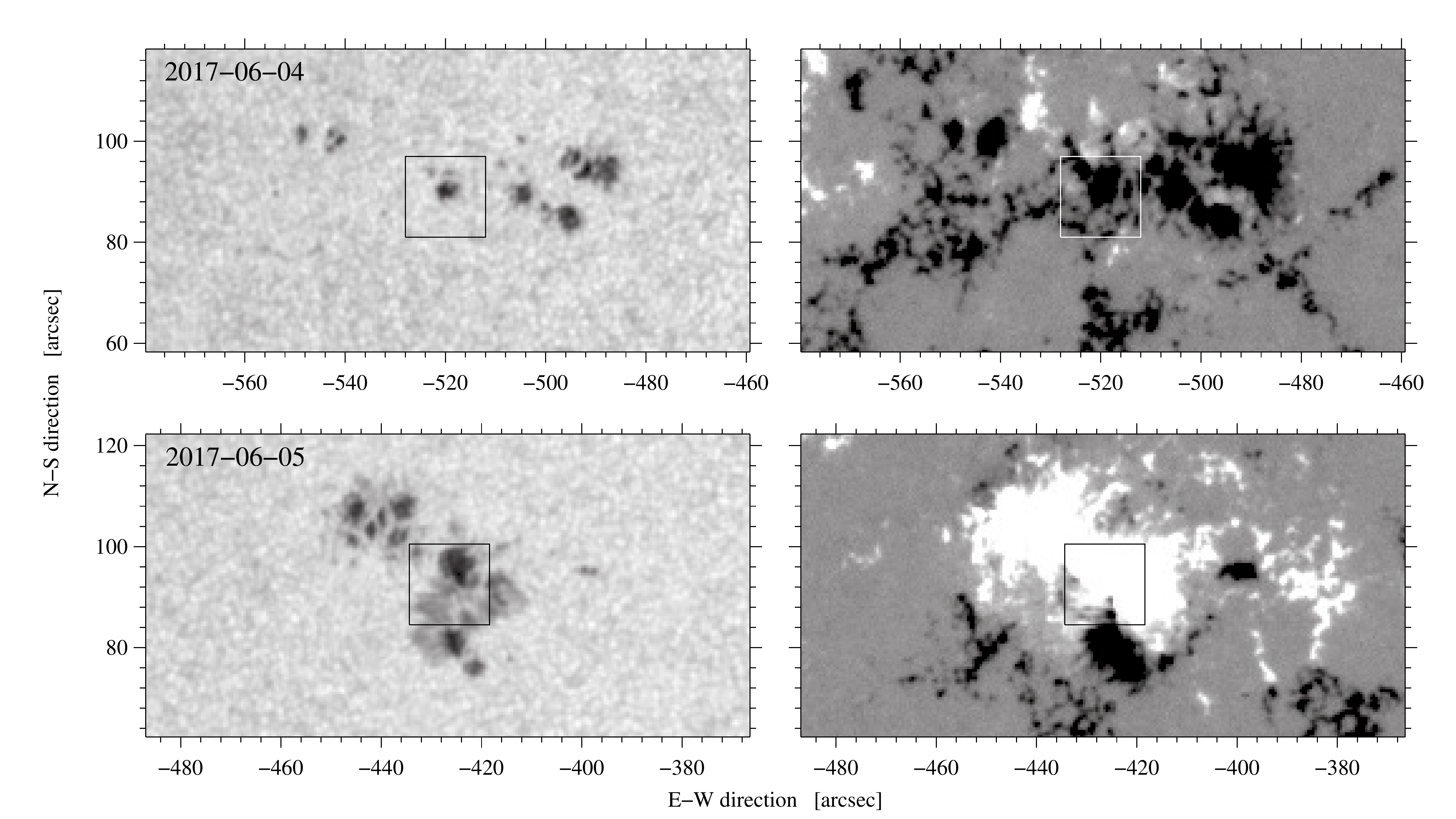}
\caption{Continuum images (\textit{left}) and line-of-sight magnetograms 
    (\textit{right}) of active region NOAA~12661 observed with HMI at 
    08:00~UT on 2017 June~4 (\textit{top}) and at 09:00~UT on 2017 June~5
    (\textit{bottom}). The axis refer to disk center coordinates. The 
    selected ROI for HiFI observations is highlighted by square boxes.} 
\label{FIG01}
\end{figure}

The observations were taken on 2017 June~4 and~5 with HiFI \citep{Denker2018c} 
at the \textit{GREGOR} solar telescope. The targets were a pore and a small 
sunspot in the leading and trailing parts of active region NOAA~12661, 
respectively. The continuum images and line-of-sight magnetograms in 
Figure~\ref{FIG01}, which were obtained with the \textit{Helioseismic and 
Magnetic Imager} \citep[HMI,][]{Scherrer2012} on board the \textit{Solar 
Dynamics Observatory} \citep[SDO,][]{Pesnell2012}, provide a general overview of 
the active region and precise pointing information. Real-time image correction 
was provided with the \textit{GREGOR Adaptive Optics System} 
\citep[GAOS,][]{Berkefeld2012}. The data reduction employed the sTools data 
processing pipeline \citep{Kuckein2017a}, which was developed by AIP's Optical 
Solar Physics group for high-resolution solar imaging and imaging 
spectropolarimetry \citep{Denker2018a}. The sTools source code and HiFI data are 
available at AIP's \textit{GREGOR} archive webpages 
(\href{https://GREGOR.aip.de}{GREGOR.aip.de}) after user registration.

Average dark and flat-field frames were applied to the HiFI data. A large number 
of flat-field frames ($n=2000$) were taken close to the center of the solar disk 
while the telescope pointing followed a circular path, thus, smearing out the 
granular pattern and on average providing a uniform illumination of the 
detector. Sensitivity and noise characteristics of sCMOS detectors differ from 
CCD technology because of active pixels, which can be considered as small 
read-out circuits. Therefore, these properties are unique for each pixel and can 
vary significantly across the detector \citep[\textit{e.g.}][]{Qiu2013}.

The two HiFI cameras are synchronized by a programmable timing unit (PTU) with 
microsecond precision. The exposure time for both detectors was either 
$t_\mathrm{exp} = 1.2$ or 1.5~ms depending on the elevation of the Sun and the 
sky brightness. Two interference filters with a full-width-at-half-maximum 
(FWHM) of about 1.1~nm selected two spectral windows, \textit{i.e.} the 
Fraunhofer G-band $\lambda$430.7~nm and a blue continuum window 
$\lambda$450.6~nm, where in the latter spectral lines are sparser than in the 
neighboring parts of the solar spectrum. Different count rates in both imaging 
channels were adjusted by inserting suitable combinations of neutral density 
filters in the brighter channel. The field-of-view (FOV) of both cameras was 
carefully aligned by inserting pinholes and a resolution target in the focal 
plane \textsf{F3} of the \textit{GREGOR} telescope \citep[see][]{Soltau2012}. 
Thus, both cameras record the same scene on the Sun at the same time under 
exactly the same seeing conditions. The light path is the same for both imaging 
channels with the exception of the final beamsplitter cube separating the two 
imaging channels. Thus, optical aberrations are virtually the same in both 
channels because aberrations introduced by the final beamsplitter cube are 
negligible.

The Imager sCMOS cameras from LaVision in G\"ottingen have $2560 \times 2160$ 
pixels but for this application only a region-of-interest (ROI) of $640 \times 
640$ pixels was selected, increasing the data acquisition rate from about 50 to 
160~Hz. The dynamic range of the images is 16~bits. Thus, the sustained writing 
speed to a RAID-0 array of eight \mbox{500-GB} solid-state-drives (SSDs) is 
250~MB~s$^{-1}$. An imaging sequence consisted typically of 50\,000 or 100\,000 
images in each channel. However, the fast acquisition rates caused intermittent 
synchronization errors between image acquisition and writing, so that the 
actually recorded number of frames is lower. Consequently, short  interruptions 
occurred when the cameras were restarted. The observing characteristics are 
summarized in Table~\ref{TAB1}, which is based on $n_4 = 451\,347$ and $n_5 = 
243\,871$ image pairs for 2017 June~4 and~5,  respectively. The total data 
volume is 1.1~TB including calibration frames. The comprehensive analysis of the 
HiFI data required several CPU weeks on regular desktop computers and multi-core 
compute servers for the image restoration. However, neither real-time processing 
nor real-time frame-selection is the primary goal of this study but can be 
accomplished for some of the image quality metrics by streamlining the 
implementation of the algorithms, by binning the data before calculating the 
metrics, by selecting smaller ROIs for the computation of the metrics, and by 
using multi-core or specialized graphical processing units (GPUs). In general, 
MFGS is not very computationally efficient considering the convolution with 
gradient operators and the median filter as a special kind of order-statistics 
filter. Some more basic but robust and efficient approaches for real-time frame 
selection were already mentioned in Section~\ref{SEC1}.

%
%

\section{Results}\label{SEC3} 

Both datasets on 2017 June~4 and~5 are included in the analysis because on the 
first day the scene on the Sun comprised a relatively simple pore and some 
granulation. This allows us to compute the granular rms-contrast 
$c_\mathrm{rms}$, which is typically used to evaluate image quality and seeing 
conditions. However, the seeing conditions were better on June~5 with the 
Fried-parameter $r_0$ reaching up to 15~cm as compared to a maximum value of 
$r_0 = 10$~cm on June~4. Thus, the seeing conditions were either very good or 
good on these two observing days but not excellent. Some of the best images from 
the \textit{GREGOR} solar telescope were presented by 
\citet{Schlichenmaier2016}. The scene on the second day comprised the central 
part of a small sunspot with umbral dots and strongly twisted penumbral 
filaments. The $r_0$-values are typically derived from the variance of the total 
mode wavefront error measured by GAOS, which records it and saves it into the 
state database of the \textit{GREGOR} Telescope Control System 
\citep[TCS,][]{Halbgewachs2012} along with other telescope information. 
Unfortunately, this recording mechanism was not available for GAOS data during 
the observations in June 2017 because of an erroneously set debugging flag in 
the control software so that only the maximum $r_0$-values from the GAOS user 
interface were at hand as written down in the observing notes.


\subsection{Median Filter Gradient Similarity}\label{SEC31}

The MFGS image quality metric was introduced to solar physics by 
\citet{Deng2015} to evaluate high-spatial resolution images. First, a 
two-dimensional median filter $F$ with a size of $3 \times 3$ pixels is applied 
to the raw image $A$, which yields an image $B = F(A)$ with diminished noise. In 
a second step, the magnitude gradients $G_A = G(A)$ and $G_B = G(B)$ are 
computed for both the raw and the filtered image. In the original 
implementation, the gradient magnitude is replaced by just a directional 
derivative $G = |G_x|$, for example, in the $x$-direction, to speed up the 
processing time. However, the magnitude gradient operator 
\begin{equation} \label{EQ1}
G = \sqrt{G_x^2 + G_y^2}
\end{equation}
can be implemented more fittingly as a convolution of the images $A$ and $B$ 
using the \citet{Scharr2007} derivative operators
\begin{equation}
G_x = \frac{1}{16}
\left[ \begin{array}{ccc}
$\phn$ -3 & 0 & $\phn$ 3\\
      -10 & 0 &       10\\
$\phn$ -3 & 0 & $\phn$ 3
\end{array} \right] \quad \mathrm{and} \quad
G_y = \frac{1}{16}
\left[ \begin{array}{ccc}
$\phm$ 3 &    $\phm$ 10 & $\phm$ 3\\
$\phm$ 0 & $\phm\phn$ 0 & $\phm$ 0\\
      -3 &          -10 &       -3\\
\end{array} \right],
\end{equation}
which contain some desirable smoothing abilities and remove much of the 
directional preference when taking just derivatives. The Scharr gradient 
operator resembles the Sobel and Prewitt edge enhancement operators but contains 
different matrix elements. The normalization factors for $G_x$ and $G_y$ can be 
omitted because they cancel when calculating the MFGS. The borders of the images 
have to be treated appropriately in both the median filter and convolution.

Finally, following \citet{Deng2015}, the MFGS value is computed as
\begin{equation}
m,\, m^\prime = \frac{2 \left(\sum G_A\right) 
    \left(\sum G_B\right)}{\left(\sum G_A\right)^2 + \left(\sum G_B\right)^2},
\end{equation}
where the symbol $\sum\ldots$ indicates the sum over all pixels in the 
gradient images. The variables $m$ and $m^\prime$ stand for the directional 
derivative $G = |G_x|$ and the magnitude gradient based on the Scharr operators 
(Equation~\ref{EQ1}), respectively. In addition, the MFGS algorithm was slightly 
altered when determining the MFGS locally, on a pixel-by-pixel basis before 
taking the sum, again using the magnitude gradient based on the Scharr operators
\begin{equation}
m^\ast = \sum \left(\frac{2 G_A G_B}{G_A^2 + G_B^2}\right),
\end{equation}
where the fraction within the parenthesis refers to a two-dimensional MFGS map. 
While the MFGS value for a single pixel is certainly not a meaningful descriptor 
of image quality, contributions to the MFGS metrics of dissimilar features in 
the FOV become more easily apparent. Taking averages over statistically 
meaningful samples, which do not have to be contiguous, can then be related to 
image quality. All image quality metrics fulfill the condition $m$, $m^\prime$, 
and $m^\ast \in (0,\ 1]$. The upper limit $m = 1$ is only reached, if both 
magnitude gradients are identical $G_A = G_B$. In the local case, $m^\ast$ is 
undefined for a very limited number of pixels, where all values within a 
3$\times$3-pixel neighborhood are identical. These pixels have to be 
appropriately replaced in the two-dimensional maps and discarded before taking 
the averages.

\begin{figure}[t]
\centering
\includegraphics[width=\textwidth]{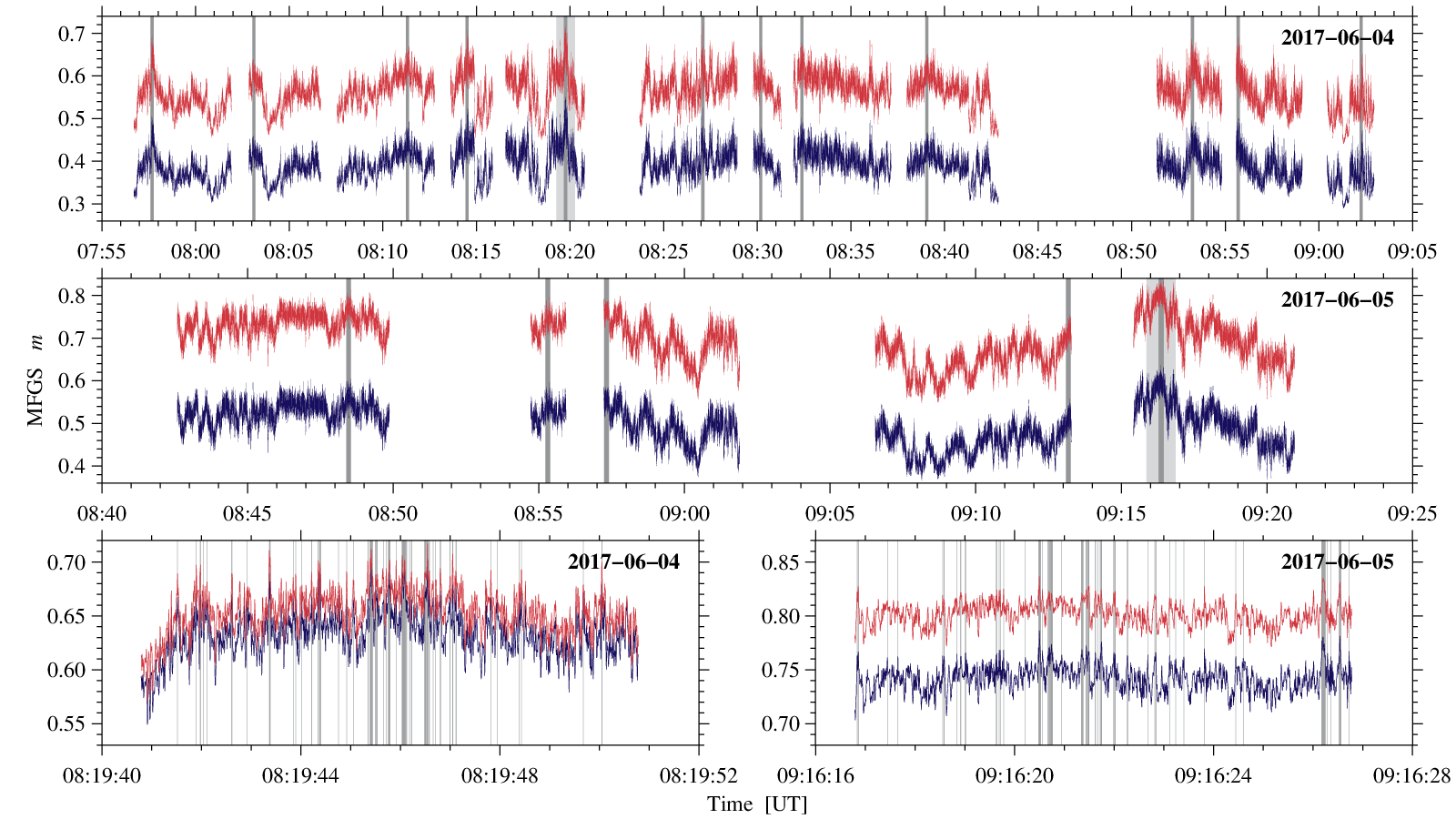}
\caption{Temporal evolution of the original MFGS value $m$ on 2017 June~4 
    (\textit{top}) and 2017 June~5 (\textit{middle}). The moments of best 
    seeing are marked by thin, dark gray rectangles indicating a 10-second 
    interval. The very best moments on each day are surrounded by broader 
    rectangles in lighter gray indicating a 1-minute time interval. The MFGS 
    records for the blue continuum images (\textit{blue}) are displaced by 0.15 
    downwards to separate them from the G-band values (\textit{red}) to
    facilitate better comparison. The MFGS values are displayed in two panels 
    (\textit{bottom}) at higher temporal resolution for the best 10-second 
    intervals on both observing days. No offsets are applied. The very best 
    $n_\mathrm{sel} = 100$ images, which are selected as input for speckle 
    masking image restoration, are marked by gray vertical lines.}
\label{FIG02}
\end{figure}


\subsection{Temporal Evolution of Seeing Conditions}\label{SEC32}

The temporal evolution of the image quality is summarized in the top and middle 
panels of Figure~\ref{FIG02} for the high-cadence blue continuum and G-band 
time-series. Computing the original MFGS value $m$ for each image yields 
profiles closely related to the prevailing seeing conditions. The image quality 
metric $m$ is highly correlated for both imaging channels and exhibits slightly 
higher values for the G-band observations. In principle, MFGS values are 
wavelength dependent and influenced by the morphological contents of an image. 
The former dependency results in larger MFGS values for longer wavelength as the 
seeing improves with increasing wavelength. However, the wavelength difference 
between the blue continuum window and the G-band spectral range is only 20~nm 
and thus negligible. The latter dependency is more relevant in this context 
because G-band images contain high-contrast bright-points caused by small-scale 
flux concentrations. These peculiarities of magnetized radiative transfer are 
exploited in proxy-magnetometry \citep{Steiner2001, Leenaarts2006}.

A total of 12 and 5 image sequences of varying length were recorded on June~4 
and~5, respectively, resulting in a 66- and a 37-minute time-series. The cause 
of the interruptions was already mentioned in Section~\ref{SEC2}. The mean MFGS 
values $m$, $m^\prime$, and $m^\ast$ are given in Table~\ref{TAB1} along with 
the rms-contrast $c_\mathrm{rms}$, which refers to a small region with 
granulation on June~4 but covers the entire ROI on June~5. Therefore, the 
contrast values are not directly comparable on both days. The implementation of 
the MFGS methods significantly affects the statistical properties and moments of 
$m$, $m^\prime$, and $m^\ast$, as is evident for the mean values of the image 
sequences. The following Section~\ref{SEC33} provides a more detailed 
correlation analysis of the various image quality metrics. On June~5, the 
original MFGS value $m$ is on average $\Delta m = 0.12$ and 0.14 higher than on 
the previous day for the blue continuum and G-band images, respectively. In 
general, the separation between the original MFGS values $m$ for blue continuum 
and G-band images increases with better seeing. The MFGS time-profiles exhibit 
substantial variations on all time-scales, \textit{i.e.} across the image 
sequences, the time-series, and the observing days. Therefore, the MFGS 
time-profiles in Figure~\ref{FIG02} impressively demonstrate the potential for 
frame selection in high-resolution solar imaging and illustrates the expected 
performance gains.

\begin{sidewaystable}[p]
\caption{Observing characteristics and image quality parameters on 2017 
    June~4 and~5.}
{\setlength\doublerulesep{0.6pt}
\begin{tabular}{ccccccccccccccc}
\toprule[1.2pt]\midrule[0.4pt]
 & & & & & & \multicolumn{4}{c}{Blue continuum} & & \multicolumn{4}{c}{G-band}\\
\cmidrule[0.4pt]{7-10}
\cmidrule[0.4pt]{12-15}
No.\rule[-6pt]{0pt}{16pt} & Date & Start & $t_\mathrm{exp}$ & $n_\mathrm{seq}$ & 
    & $m$ & $m^\prime$ & $m^\ast$ & $c_\mathrm{rms}$ & & $m$ & 
    $m^\prime$ & $m^\ast$ & $c_\mathrm{rms}$\\
\midrule[0.4pt]
\phn 1 & 2017--06--04 & 07:56:42~UT & 1.5~ms & \phn 50\,000 & & 0.521 & 0.855 & 0.745 & \phn 1.95\% & & 0.541 & 0.874 & 0.759 & \phn 2.87\% \\ 
\phn 2 & 2017--06--04 & 08:02:51~UT & 1.5~ms & \phn 36\,702 & & 0.529 & 0.865 & 0.753 & \phn 2.06\% & & 0.550 & 0.884 & 0.767 & \phn 3.09\% \\ 
\phn 3 & 2017--06--04 & 08:07:33~UT & 1.5~ms & \phn 50\,000 & & 0.542 & 0.877 & 0.761 & \phn 2.25\% & & 0.569 & 0.898 & 0.778 & \phn 3.75\% \\ 
\phn 4 & 2017--06--04 & 08:13:37~UT & 1.5~ms & \phn 21\,554 & & 0.542 & 0.873 & 0.761 & \phn 2.58\% & & 0.564 & 0.890 & 0.774 & \phn 3.65\% \\ 
\phn 5 & 2017--06--04 & 08:16:34~UT & 1.5~ms & \phn 40\,514 & & 0.549 & 0.878 & 0.764 & \phn 2.65\% & & 0.567 & 0.891 & 0.774 & \phn 3.43\% \\ 
\phn 6 & 2017--06--04 & 08:23:43~UT & 1.2~ms & \phn 50\,000 & & 0.546 & 0.879 & 0.763 & \phn 2.31\% & & 0.572 & 0.898 & 0.776 & \phn 3.00\% \\ 
\phn 7 & 2017--06--04 & 08:29:47~UT & 1.2~ms & \phn 14\,321 & & 0.544 & 0.877 & 0.761 & \phn 2.17\% & & 0.570 & 0.895 & 0.773 & \phn 2.87\% \\ 
\phn 8 & 2017--06--04 & 08:31:56~UT & 1.2~ms & \phn 50\,000 & & 0.553 & 0.886 & 0.769 & \phn 2.32\% & & 0.580 & 0.905 & 0.783 & \phn 3.66\% \\ 
\phn 9 & 2017--06--04 & 08:37:59~UT & 1.2~ms & \phn 46\,940 & & 0.533 & 0.865 & 0.753 & \phn 2.30\% & & 0.558 & 0.886 & 0.767 & \phn 3.62\% \\ 
    10 & 2017--06--04 & 08:51:21~UT & 1.2~ms & \phn 33\,344 & & 0.545 & 0.879 & 0.764 & \phn 2.50\% & & 0.571 & 0.898 & 0.777 & \phn 3.45\% \\ 
    11 & 2017--06--04 & 08:55:34~UT & 1.2~ms & \phn 33\,918 & & 0.529 & 0.864 & 0.752 & \phn 2.16\% & & 0.552 & 0.885 & 0.767 & \phn 3.20\% \\ 
    12 & 2017--06--04 & 09:00:26~UT & 1.2~ms & \phn 24\,054 & & 0.500 & 0.829 & 0.729 & \phn 1.90\% & & 0.515 & 0.847 & 0.740 & \phn 2.88\% \\ 
\cmidrule[0.4pt]{5-5}\cmidrule[0.4pt]{7-10}\cmidrule[0.4pt]{12-15}
       &              &             &        &     451\,347 & & 0.537 & 0.870 & 0.757 & \phn 2.26\% & & 0.560 & 0.889 & 0.770 & \phn 3.32\% \\
\midrule[0.4pt]
\phn 1 & 2017--06--05 & 08:42:35~UT & 1.2~ms & \phn 69\,911 & & 0.676 & 0.959 & 0.840 &     16.05\% & & 0.733 & 0.973 & 0.859 &     18.20\% \\
\phn 2 & 2017--06--05 & 08:54:44~UT & 1.2~ms & \phn 11\,644 & & 0.672 & 0.957 & 0.837 &     15.61\% & & 0.728 & 0.971 & 0.856 &     17.74\% \\
\phn 3 & 2017--06--05 & 08:57:14~UT & 1.2~ms & \phn 44\,689 & & 0.639 & 0.943 & 0.822 &     12.61\% & & 0.691 & 0.961 & 0.842 &     14.53\% \\
\phn 4 & 2017--06--05 & 09:06:33~UT & 1.2~ms & \phn 64\,612 & & 0.603 & 0.926 & 0.804 &     10.27\% & & 0.649 & 0.946 & 0.824 &     11.99\% \\
\phn 5 & 2017--06--05 & 09:15:25~UT & 1.2~ms & \phn 53\,015 & & 0.653 & 0.948 & 0.828 &     13.08\% & & 0.709 & 0.964 & 0.848 &     15.08\% \\
\cmidrule[0.4pt]{5-5}\cmidrule[0.4pt]{7-10}\cmidrule[0.4pt]{12-15}
       &              &            &         &     243\,871 & & 0.644 & 0.945 & 0.824 &     13.22\% & & 0.698 & 0.961 & 0.844 &     15.18\% \\
\midrule[0.4pt]\bottomrule[1.2pt]
\end{tabular}
\label{TAB1}
}
\end{sidewaystable}

The two panels in the bottom row of Figure~\ref{FIG02} display 10-second MFGS 
time-profiles during the moments of best seeing (taken from image sequences 
No.~5 on both observing days). These moments are also marked by thin, dark gray 
rectangles in the top and middle panels for each image sequence. The very best 
moments are highlighted by broader rectangles in light gray  corresponding to a 
1-minute time interval. Out of the $n_\mathrm{set} = 1600$ images per 10-second 
time interval only $n_\mathrm{sel} = 100$ images are selected for subsequent 
image restoration. These instants are indicated by gray vertical lines, which 
cover the entire 10-second time interval. However, both MFGS time-profiles on 
June~4 and~5 exhibit a tendency of clustering, \textit{i.e.} some 5\,--\,10 high 
values occur almost consecutively. These clusters of a few high MFGS values can 
likely be expanded by increasing the data acquisition rate by a factor 5\,--\,6 
considering the 1-millisecond exposure times, if faster detectors become 
available. As a result, the time interval can be shortened to acquire images 
with statistically independent wavefront aberrations required for image 
restoration. The current data acquisition rate of $f_\mathrm{acq} = 160$~Hz is 
still insufficient to reach the point, where wavefront aberrations are constant 
in consecutive images. The above frame selection parameters are similar to the 
current HiFI default observing sequence, which uses $n_\mathrm{sel} = 100$ and 
$n_\mathrm{set} = 500$ for a 10-second time interval. The later parameter is 
lower because full-format frames with $2 \times 5$ megapixels are recorded and 
not just small ROIs. These settings can of course be adapted to a specific 
science case -- as in this study. Since frame selection is performed after image 
acquisition, it is usually known, if interesting transient events occurred (for 
example flares). In this case, or if the data were taken under excellent seeing 
conditions, all images will be kept to retain the highest temporal evolution or 
to zoom in on a particular time period.

\begin{figure}[t]
\centering
\includegraphics[width=\textwidth]{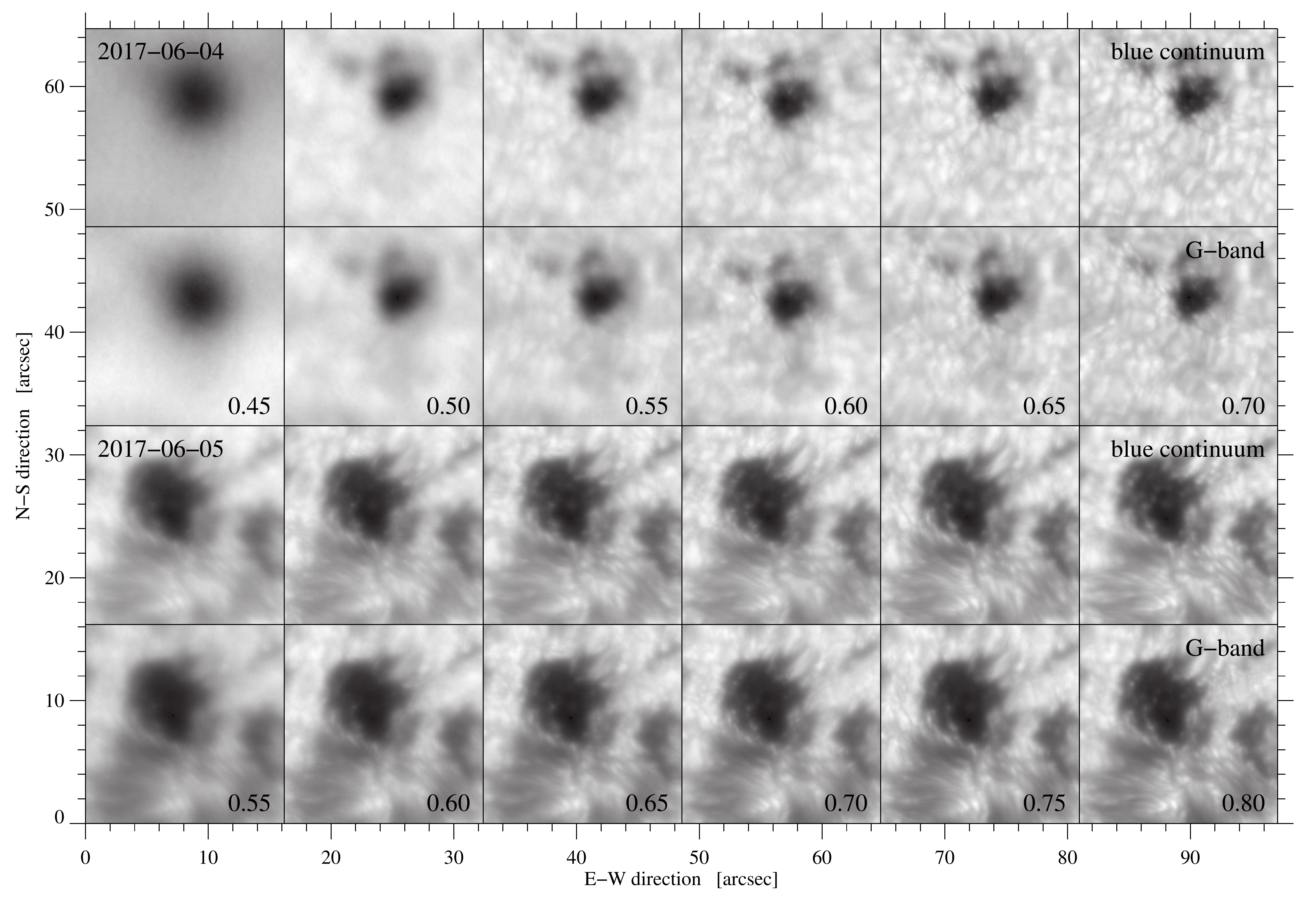}
\caption{Blue continuum and G-band images of a pore (\textit{top two rows})
    and a small sunspot (\textit{bottom two rows}) in active region NOAA~12661 
    observed with HiFI on 2017 June~4 and~5, respectively. The images are
    normalized with respect to the quiet-Sun intensity $I_0$ and displayed in 
    the range $I/ I_0 \in [0.4,\ 1.1]$ and $I/ I_0 \in [0.5,\ 1.2]$, 
    respectively. The original MFGS value $m$ is given in the bottom-right 
    corner for the G-band images.} 
\label{FIG03}
\end{figure}

Visually judging and ranking image quality for high-resolution images is a 
relatively simple task and requires only minimal training 
\citep[\textit{e.g.}][]{Zirin1988b}. However, the huge data volume renders this 
task impracticable and even impossible for high-cadence and high-resolution 
solar images. The observations on June~4 and~5 covered a wide range of seeing 
conditions. Figure~\ref{FIG03} relates the visual impression of image quality to 
the quantitative MFGS value $m$. This value is given just for the G-band images 
in equidistant steps of $\Delta m = 0.05$. On both observing days, image 
sequences No.~5 were used to find G-band images, where the MFGS values closely 
match the thus defined thresholds ($m = 0.45,\ 0.50,\ 0.55, \ldots$). The 
corresponding values for the blue continuum images are on average lower by 
$\Delta m = 0.02$\,--\,0.05 but follow the same trend. Below the threshold of $m 
= 0.5$ for G-band images, solar fine structure like the granulation pattern 
completely vanishes, and only strong features like dark pores persevere. A MFGS 
threshold of $m > 0.65$ for G-band images is a good choice for image restoration 
using the above frame selection criterion (best $n_\mathrm{sel} = 100$ images 
out of $n_\mathrm{set} = 1600$ images in a 10-second time interval). However, on 
June~4, this criterion is only fulfilled for $N_\mathrm{set} = 9$ out of 277 
possible 10-second intervals. On June~5, the seeing conditions are much 
improved, so that the criterion is surpassed in $N_\mathrm{set} = 143$ out of 
150 cases. An even stricter criterion of $m = 0.75$ for G-band images still 
yields about one third of all possible 10-second intervals.

The following sections illustrate the results of the analysis using subsets 
(June~4 or~5 as well as G-band or blue continuum images) of the image sequences 
for reasons of conciseness but the findings are based on the complete dataset. 
Notwithstanding, major differences are explicitly noted.


\subsection{Correlations between Image Quality Metrics}\label{SEC33}

In this study, four parameters are used to describe image quality and in turn 
the prevailing seeing conditions, \textit{i.e.} the granular rms-contrast 
$c_\mathrm{rms}$ and three implementations of MFGS $m$, $m^\prime$, and 
$m^\ast$. The goal of the correlation analysis is to determine relationships 
among these parameters and to identify differences when comparing blue continuum 
and G-band images. 

The granular rms-contrast $c_\mathrm{rms}$ is commonly used to characterize 
seeing conditions relying on the uniform and isotropic properties of 
granulation. However, granulation exhibits a distinct center-to-limb variation 
\citep[\textit{e.g.}][]{Wilken1997, Carlsson2004}, experiences geometric 
foreshortening close to the solar limb, and depends on the observed wavelength 
\citep[\textit{e.g.}][]{WedemeyerBoehm2009b}. In addition, telescope optics and 
instrumental straylight have a significant impact on the granular contrast. Only 
the images observed on June~4 contain a considerable fraction of granulation. 
Thus, the correlation analysis is based on these data. Computing MFGS values for 
the region with granulation or the entire FOV only results in minor differences, 
thus MFGS values are based on the full FOV. Note that the rms-contrast for the 
sunspot with rudimentary penumbra is an order of magnitude higher than that of 
granulation, which motivates the search for image quality metrics with weak or 
negligible dependency on the observed scene on the solar surface.

\begin{figure}
\centering
\includegraphics[width=\textwidth]{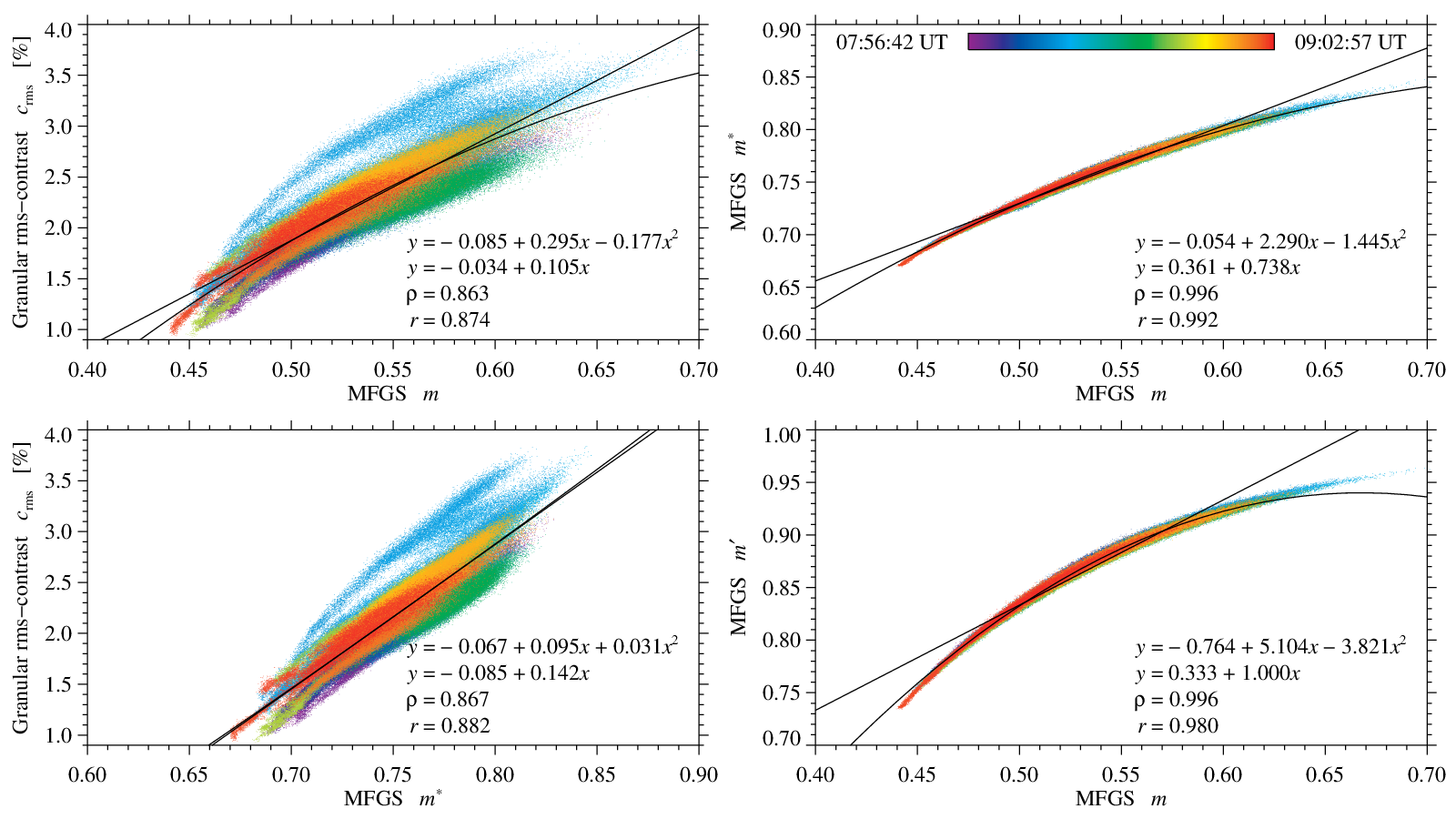}
\caption{Scatter plots of granular rms-contrast $c_\mathrm{rms}$ and various 
    MFGS implementations $m$, $m^\prime$, and $m^\ast$ for the 2017 June~4 
    G-band images. The observing time is color coded according to the scale 
    bar in the top-right panel. However, some of the earlier data points are 
    covered by those obtained later. The black lines and curves represent
    linear and parabolic models, respectively, where the corresponding
    mathematical expressions are given in the bottom-right corner of the 
    panels along with Pearson's linear correlation coefficient $r$ and
    Spearman's rank-order correlation coefficient $\rho$.}
\label{FIG04}
\end{figure}

The two left panels of Figure~\ref{FIG04} compare the rms-granular contrast 
$c_\mathrm{rms}$ with the two image quality metrics $m$ and $m^\ast$ for blue 
continuum images. Both scatter plots show a broad distribution around the linear 
regression line. Even a $2^\mathrm{nd}$ order polynomial does not improve the 
regression. Visual inspection of the two scatter plots hints at a more linear 
dependency between $c_\mathrm{rms}$ and the local MFGS implementation $m^\ast$ 
as compared to the original MFGS procedure $m$ by \citet{Deng2015}. These 
authors already mentioned that the (granular) rms-contrast has an inferior 
performance compared to MFGS methods. On the other side, the region with 
granulation is relatively small containing only a few tens of granules. 
Therefore, the temporal evolution of granules potentially affects the 
rms-contrast because the sample is too small violating the assumption that 
granulation is on average uniform and isotropic.

The two right panels of Figure~\ref{FIG04} compare the two alternative MFGS 
implementations $m^\prime$ and $m^\ast$ to the original approach for the MFGS 
metric $m$ by \cite{Deng2015}. The correlations in these scatter plots are much 
tighter, though not linear. However, the trend in the scatter plots is 
monotonically increasing, and a $2^\mathrm{nd}$ order polynomial fit already 
provides satisfactory regression results, capturing the functional dependency. 
Interestingly, just switching from the directional derivative to the magnitude 
gradient based on Scharr operators (see Section~\ref{SEC31}) deviates from a 
linear model and changes the range of MFGS values considerably. Note that the 
range of values is kept similar for abscissae and ordinates in the panels of 
Figure~\ref{FIG04} facilitating easier visual comparison. The deviation from a 
linear model is even stronger for the local MFGS values $m^\ast$ but the 
monotonic trend in the scatter plot still allows to establish a strict 
functional dependency between $m$ and $m^\ast$. \textit{Ab initio}, this is not 
anticipated but justifies determining the MFGS locally, on a pixel-by-pixel 
basis -- thus gaining access to the structure-dependency of MFGS within the 
observed FOV.

\begin{figure}
\centering
\includegraphics[width=0.65\textwidth]{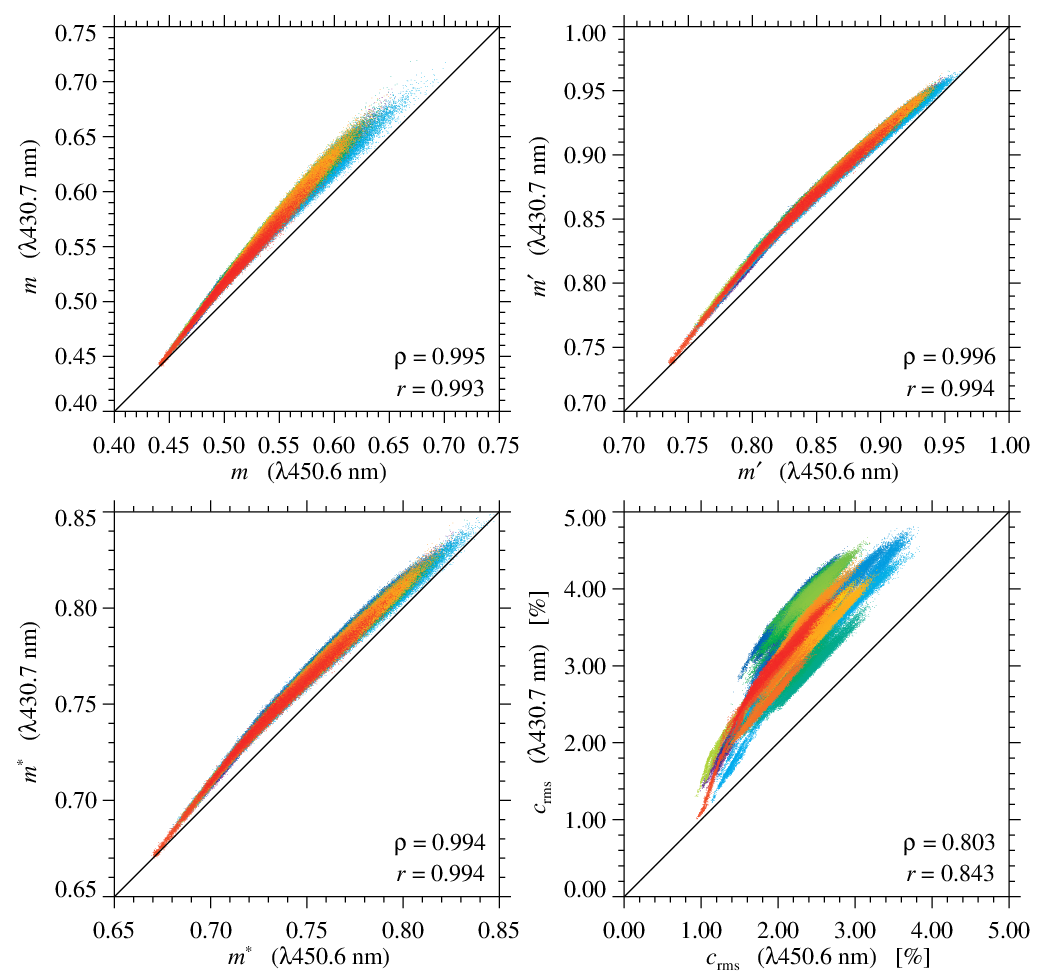}
\caption{Scatter plots of granular rms-contrast $c_\mathrm{rms}$ and various 
    MFGS implementations $m$, $m^\prime$, and $m^\ast$ comparing the same
    parameters for blue continuum and G-band images observed on 2017 June~4. 
    The color code is the same as in Figure~\ref{FIG04}, and Pearson's linear
    correlation coefficient $r$ and Spearman's rank-order correlation 
    coefficient $\rho$ are given in the bottom-right corner of the panels.}
\label{FIG05}
\end{figure}

Even though observed at neighboring wavelengths, the morphological differences 
between blue continuum and G-band images lead to noticeable variations in the 
four image quality respectively seeing parameters $c_\mathrm{rms}$, $m$, 
$m^\prime$, and $m^\ast$. The parameters for blue continuum and G-band images do 
not coincide with the line of equality in the scatter plots of 
Figure~\ref{FIG05}. The G-band values are always above the identity line. For 
$m^\prime$ and $m^\ast$, the values approach the identity line only for the 
lowest and highest values, indicating that either the fine structure is 
completely washed out in bad seeing conditions or that the seeing is so good 
that the discriminatory power of both MFGS implementations becomes very similar. 
The former effect is also observed for $c_\mathrm{rms}$ and $m$. However, under 
good seeing conditions, the trendline becomes more parallel to the line of 
equality. As in Figure~\ref{FIG04}, the values of the granular rms-contrast are 
widely dispersed in the scatter plot shown in the bottom-right panel of 
Figure~\ref{FIG05}. These trends are also reflected in Pearson's linear 
correlation coefficients $r$ and Spearman's rank-order correlation coefficients 
$\rho$, given in the bottom-left corner of the plot panels, which are much lower 
for the rms-granular contrast $c_\mathrm{rms}$ than for the MFGS parameters $m$, 
$m^\prime$, and $m^\ast$.


\subsection{Impact of Image Acquisition Rate}\label{SEC34}

The high-cadence time-series of blue continuum and G-band images allow us to 
evaluate the impact of the image acquisition rate $f_\mathrm{acq}$ on the image 
quality metrics $m$ after selecting the best images for image restoration. For 
this numerical experiment, $n_\mathrm{sel} = 100$ images are selected in a time 
interval $\Delta t_\mathrm{set} = 10$~s, which contains a set of $n_\mathrm{set} 
= 1600$ images. Thus, the total number of consecutive sets is $N_\mathrm{set} = 
427$ on both observing days. The MFGS values $m_{i,j}$ refer to individual 
images in a set, whereas $m_i$ relates to all MFGS values in a given set, and 
$\overline{m}_i$ is the mean value of all $n_\mathrm{set} = 1600$ images in a 
set. The improvement in the image quality metrics $m$ is given in percent by the 
expression
\begin{equation}
q(f_\mathrm{acq}) = \frac{100\%}{N_\mathrm{set}} \sum_{i=1}^{N_\mathrm{set}} 
    \frac{\overline{m}_i(f_\mathrm{acq}) - \overline{m}_i}{\overline{m}_i}
    \quad \mathrm{with}
\end{equation}
\begin{equation}
\overline{m}_i = \frac{1}{n_\mathrm{set}} \sum_{j=1}^{n_\mathrm{set}}
    m_{i,j} \quad \mathrm{with} \quad \overline{m}_i 
    (f_\mathrm{acq}) = \frac{1}{n_\mathrm{sel}} 
    \sum_{j=1}^{n_\mathrm{sel}} \mathrm{best}\left(n_\mathrm{acq},\ m_{i,j}
    \right), 
\end{equation}
where the function $\mathrm{best}(\ldots)$ selects the best $n_\mathrm{sel} = 
100$ images from a subset of $n_\mathrm{acq} = 100$, 150, 200, \ldots, 
$n_\mathrm{set}$ images, which are evenly spaced within a set of 
$n_\mathrm{set}$ images. Thus, $n_\mathrm{acq}$ corresponds to the images 
acquisition rate $f_\mathrm{acq}$. Since $n_\mathrm{set}\ \mathrm{mod}\ 
n_\mathrm{acq} \neq 0$ in many cases, the indices of the $n_\mathrm{acq}$ images 
were first computed as real numbers before casting them with the floor function 
to integers. 

\begin{figure}
\centering
\includegraphics[width=\textwidth]{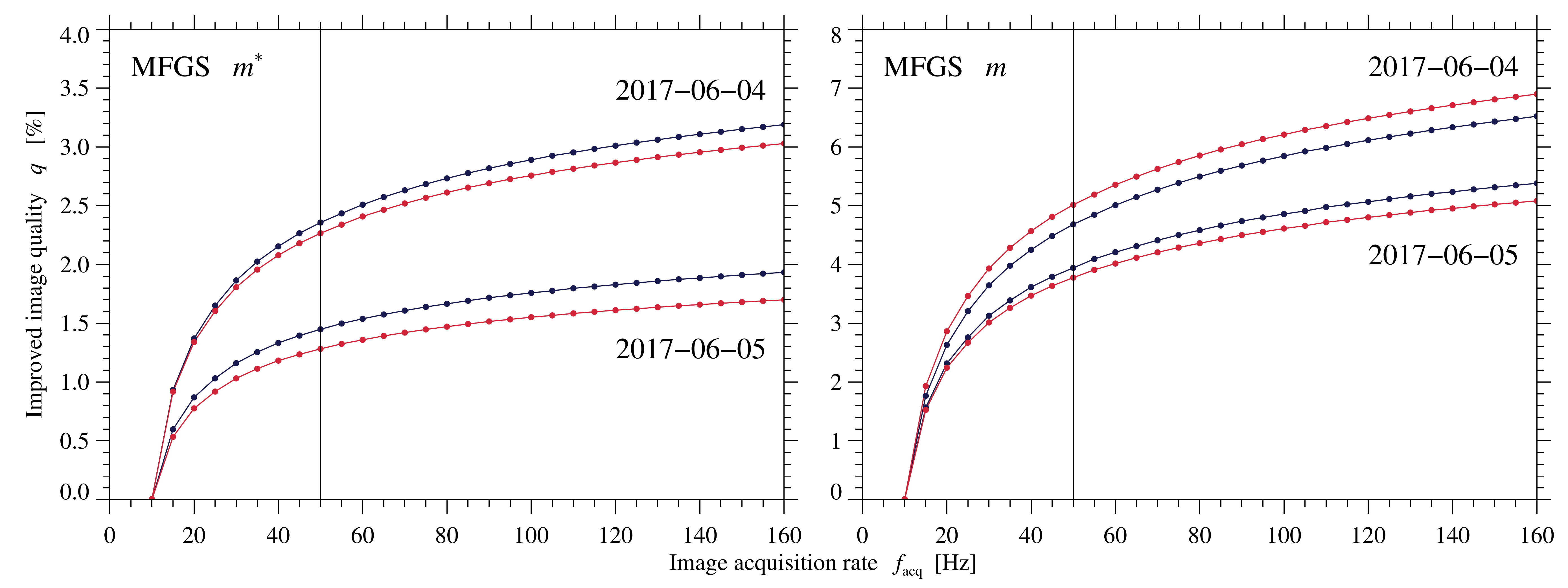}
\caption{Improvement of the image quality $q$ as a function of the image
    acquisition rate $f_\mathrm{acq}$ when using frame selection 
    ($n_\mathrm{sel} = 100$ images). The quality parameters $q(f_\mathrm{acq})$ 
    are computed based on the local MFGS $m^\ast$ (\textit{left}) and on the 
    original implementation of the MFGS $m$ (\textit{right}), which are given 
    for both observing days and the blue continuum (\textit{blue}) and G-band 
    (\textit{red}) images. The bullets mark the image quality measurements at 
    intervals of 10~Hz. The vertical lines at $f_\mathrm{acq} = 50$~Hz refer to 
    the data acquisition rate of full-format sCMOS frames.}
\label{FIG06}
\end{figure}

The results from this numerical experiment are presented in Figure~\ref{FIG06} 
for both observing days, both the blue continuum and G-band images, and both 
image quality metrics $m^\ast$ and $m$. At the data acquisition rate 
$f_\mathrm{acq} = 10$~Hz, the image quality basically remains the same, as no 
image selection is carried out. Slight differences only arise from the coarser 
sampling of the time interval $\Delta t_\mathrm{set} = 10$~s with 
$n_\mathrm{sel} = 100$ instead of $n_\mathrm{set} = 1600$ images. At higher 
frequencies, the image quality improves monotonically, with most of the gain at 
frequencies below $f_\mathrm{acq} = 50$~Hz, which is, by coincidence, also the 
data acquisition rate of the HiFI sCMOS sensors when reading out full-format 
frames. Even though higher frame rates are desirable, off-the-shelf and 
state-of-the-art sCMOS detectors are already a very good choice for 
high-resolution solar imaging.

By visual comparison, the improvement of image quality based on the MFGS value 
$m$ is about twice that of $m^\ast$. However, considering that the range covered 
by both image quality metrics also differs, the absolute improvement in  percent 
is not a decisive factor. The slopes of $q(f_\mathrm{acq})$ for higher image 
acquisition rates, however, indicate that the original MFGS $m$ is more 
sensitive to changes of the image quality in this region. Typically, the blue 
continuum images experience a larger improvement of the image quality, with 
exception of the MFGS values $m$ on June~4. A likely explanation is the 
directional derivative in the original implementation by \citet{Deng2015}. Using 
the gradient magnitude based on Scharr operators but the same summation scheme 
as in \citet{Deng2015} already delivers MFGS values $m^\prime$ very similar to 
those in the left panel of Figure~\ref{FIG06} for the local MFGS $m^\ast$. If 
the seeing conditions are moderately good, as they were on June~4, the marginal 
benefits of frame selection are higher than on days, when the seeing is very 
good, \textit{e.g.} on June~5. As a result, more image sets reach the threshold 
where image restoration becomes feasible, which significantly improves the 
temporal coverage during an observing campaign.


\subsection{Speckle Masking Image Restoration}\label{SEC35}

Obtaining sets of high-quality images is the goal of frame selection and input 
for \textit{post-facto} image restoration. The standard method to restore HiFI 
images is the triple correlation or speckle masking technique 
\citep{Lohmann1983, Weigelt1983, deBoer1993, vonderLuehe1993}. The selected 
images are restored with the Kiepenheuer Institute Speckle Interferometry 
Package \citep[KISIP,][]{vonderLuehe1993, Woeger2008a, Woeger2008b}. The 
algorithm includes an estimation of the long-exposure and speckle transfer 
functions and a field-dependent calibration of the Fourier amplitudes for the 
reconstructed sub-images \citep{Woeger2007, Woeger2010}. The restored sub-images 
have a size of $256 \times 256$ pixels, which corresponds to $6.5\arcsec \times 
6.5\arcsec$ and is somewhat larger than the size of the isoplanatic patch under 
daytime seeing conditions \citep{Roddier1982, Irbah1993}. However, taking into 
account the apodization of the sub-images, their useful size reduces to about 
$200 \times 200$ pixels or about $5\arcsec \times 5\arcsec$. Finally, the 
scattered light was corrected, which is introduced by Earth's atmosphere and 
imperfect telescope/instrument optics \citep[see \textit{e.g.}][and references 
therein]{BelloGonzalez2008}, by deconvolution with an appropriate point-spread 
function. 

The previous sections implicitly assumed that for HiFI observations 
$n_\mathrm{sel} = 100$ and a 10-second time interval are appropriate choices for 
speckle masking image restoration. The underlying assumption for the latter 
parameter is that features are typically not moving faster than 2~km~s$^{-1}$, 
which is already a concession based on the cost-benefit argument that only few 
pixels exhibit higher velocities. Studying transient events, thus requires 
different observing parameters. The optimal number of input images for image 
restoration depends on many factors, most notably on the image restoration 
algorithm and the seeing conditions. For example, multi-frame blind 
deconvolution \citep[MFBD,][]{Loefdahl2002, vanNoort2005} require just a few 
images of very good quality. To determine the proper number of input images, the 
best set in sequence No.~5 on June~5 was chosen for the analysis. Image quality 
metrics and photometric error were computed for $n_\mathrm{sel} = 10,\ 20,\ 
\ldots,\ 100,\ 200,\ \ldots,\ 1600$ input images. The photometric error refers 
to the standard deviation of the intensity difference between restored image and 
the image restoration using all $n_\mathrm{set} = 1600$ images, divided by the 
mean intensity the restored reference image. This assumes that all images 
contribute to an improved image quality, which is valid as long as all images 
contain some diffraction-limited information. Obviously, if the seeing 
conditions are only mediocre, this assumption is violated, which might lead to a 
loss of image quality in the restored image. In the computation of image quality 
metrics and photometric error only the inner $4 \times 4$ restored isoplanatic 
patches were considered, after removal of the apodization borders.

\begin{figure}
\centering
\includegraphics[width=\textwidth]{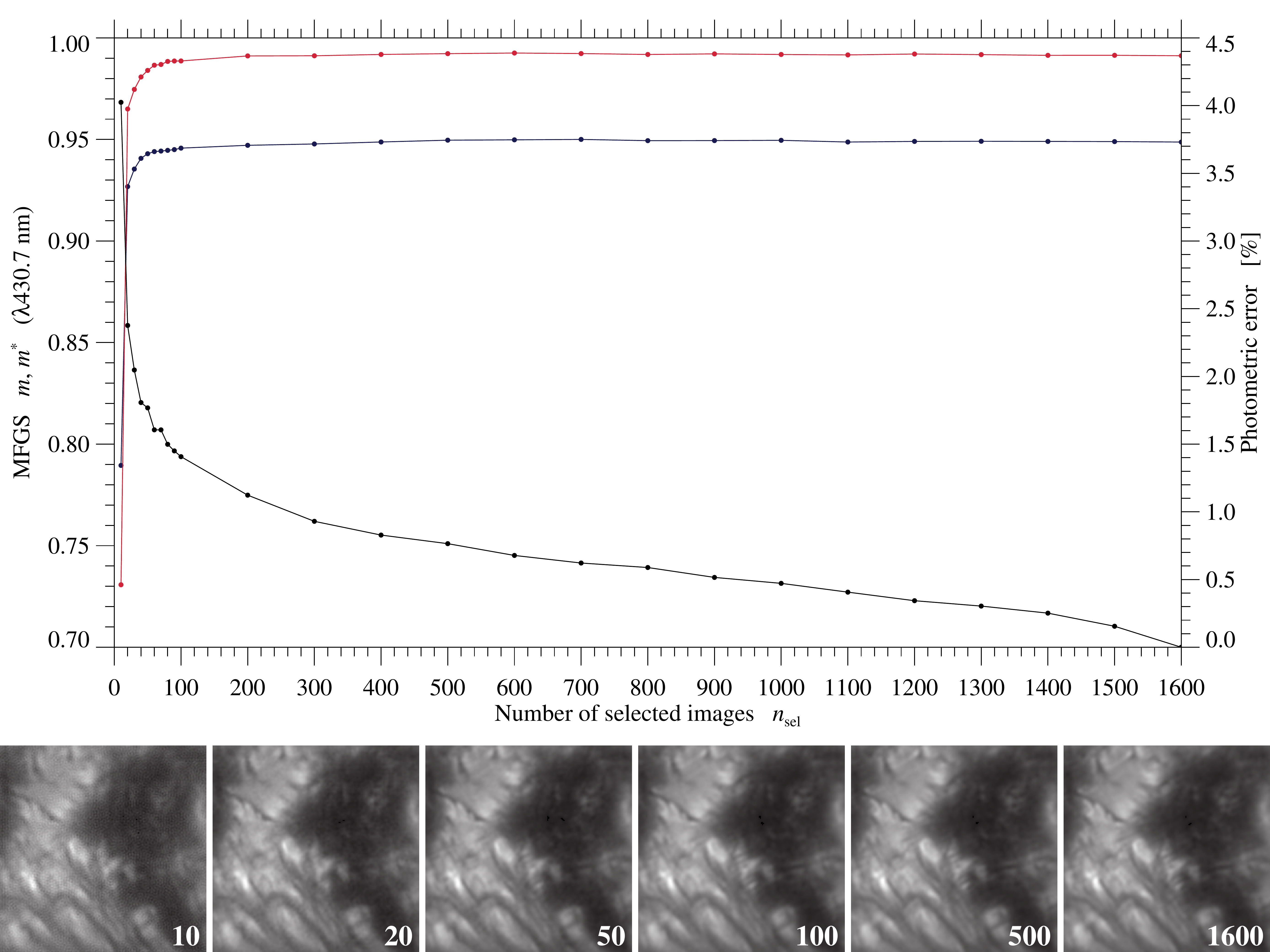}
\caption{Image quality metrics MFGS $m$ (\textit{red}) and $m^\ast$ 
    (\textit{blue}) as a function of the number of frame-selected input
    images $n_\mathrm{sel}$ that were used for the speckle masking image 
    restoration. The sampling is ten times finer for $n_\mathrm{sel} < 100$.
    The scale on the right side of the plot panel refers to the photometric 
    error (\textit{black}) using the restored image for $n_\mathrm{sel} = 1600$
    as a reference. The restored $5\arcsec \times 5\arcsec$-patches at the
    bottom provide a visual impression of the image quality as a function of
     $n_\mathrm{sel}$, which is given in the lower-right corner of each patch.}
\label{FIG07}
\end{figure}

The image quality metrics MFGS $m$ and $m^\ast$ are given in Figure~\ref{FIG07} 
as a function of the number of input images $n_\mathrm{sel}$. The behaviour of 
the metrics $m^\ast$ and $m^\prime$ is very similar so that the latter was 
omitted to avoid clutter. The original MFGS implementation $m$ is always larger 
than the locally computed $m^\ast$. Nonetheless, both metrics show the same 
trend, \textit{i.e.} a steep increase for low values of $n_\mathrm{sel}$, 
levelling out at $n_\mathrm{sel} \approx 100$. The photometric error drops below 
1\% for $n_\mathrm{sel} \approx 250$. Science cases based on high-resolution 
images typically require that morphological changes can be clearly identified, 
\textit{i.e.} a good threshold for $n_\mathrm{sel}$ is reached when images a 
visually identical. The human eye can adapt to a wide range of intensities by 
brightness adaptation. However, discriminating between distinct gray levels at 
the same time is much more limited, and in this respect the dynamic range of the 
human eye is only 6\,--7 bits \citep{Gonzalez2002}. Therefore, $n_\mathrm{sel} = 
100$ is a good choice considering the marginal benefits. This is also 
illustrated in the bottom row of Figure~\ref{FIG07}, where restored isoplanatic 
patches are displayed as a function of $n_\mathrm{sel}$. For $n_\mathrm{sel} 
\leq 50$, noise still dominates the restored patches, including even ``ringing'' 
artifacts for $n_\mathrm{sel} = 10$. This indicates that insufficient 
high-spatial frequency information was available in the Fourier domain. However, 
even for the case of just 10 input images, nearly diffraction-limited 
information is recovered, \textit{i.e.} the phase recovery is already very good. 
On the other hand, photometric accuracy, which is mainly encoded in the Fourier 
amplitudes, scales roughly with the square root of the number of detected 
photons, \textit{i.e.} the number of input images $n_\mathrm{sel}$. This weak 
dependence results only in an improved photometry by a factor of four when 
increasing $n_\mathrm{sel} = 100$ to 1600. The improvements are most clearly 
seen in bright-points superposed on a dark background, like umbral dots, where 
improved photometry removes much of the surrounding diffuse halo. An additional 
benefit of using more images in the restoration process is the better 
suppression of fixed-pattern noise. A more thorough analysis and discussion of 
photometric precision in speckle masking imaging is provided by \citet{Peck2017} 
with an emphasis on AO-corrected images from large-aperture solar telescopes.

\begin{figure}
\centering
\includegraphics[width=\textwidth]{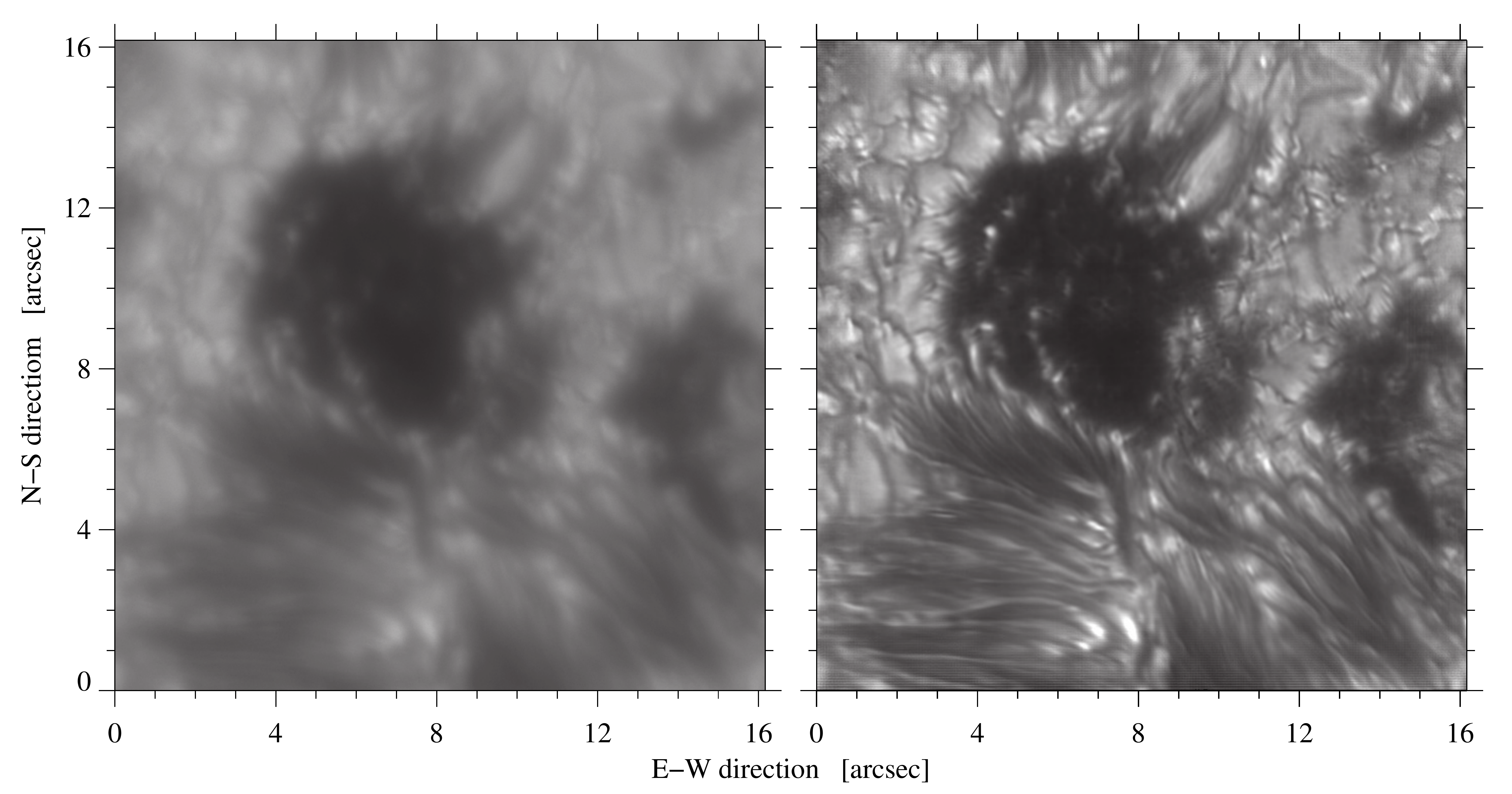}
\caption{Best G-band image with $m = 0.84$ (\textit{left}) and matching
    speckle-restored image with $m = 0.99$ (\textit{right}) of a small sunspot 
    with rudimentary penumbra in the trailing part of active region NOAA~12661 
    observed at 09:16:17~UT on 2017 June~5. Both images were scaled in the 
    range $I / I_0 \in [0.1,\ 1.5]$, which is adapted to the intensity range of 
    the restored G-band image.}
\label{FIG08}
\end{figure}

Figure~\ref{FIG08} compares the best image of the best set in sequence No.~5 
observed on June~5 with the corresponding speckle-restored and scattered-light 
corrected image, which shows a clear enhancement of the contrast and of the 
fine-structure contents. The horizontal fringing in the upper and lower parts of 
the restored G-band image is an artifact caused by the readout registers at the 
top and bottom of the sCMOS detectors, which create a fixed noise pattern with 
amplitudes and spatial frequencies close to those of solar fine structure. 
Consequently, these artifacts survive the image restoration and might even be 
enhanced in the process.

Purposefully, both images are scaled according to the intensity range of the 
restored G-band image to demonstrate that speckle masking image restoration 
affects both amplitudes and phases in the Fourier domain. G-band bright-points 
and substructure of umbral dots are clearly visible with fine structure close to 
the diffraction limit of the \textit{GREGOR} solar telescope. In addition, some 
penumbral filaments exhibit the distinctive dark cores first reported by 
\citet{Scharmer2002}, which are extremely thin with a width of 0.2\arcsec\ or 
even less. Interestingly, some penumbral filaments appear almost at right angles 
to each other and almost tangential to the umbral core, which is indicative of 
highly sheared and twisted magnetic field lines. 

The two-dimensional power spectrum of the speckle-restored image was averaged in 
the azimuthal direction to yield a one-dimensional power spectrum, which 
facilitates estimating the spatial cut-off frequency. This frequency is 15\,--16 
arcsec$^{-1}$ and corresponds to a structure size of 0.065\arcsec. The exact 
value depends on the definition of the cut-off frequency and the settings of the 
noise filter. This value was also confirmed using MFBD for image restoration, 
where the cut-off frequency is slightly smaller, \textit{i.e.} 14\,--15 
arcsec$^{-1}$, which corresponds to 0.069\arcsec. The diffraction limit for the 
G-band images is $\lambda / D = 0.062\arcsec$, where $\lambda$ is the observed 
wavelength, and $D$ refers to the diameter of the telescope aperture. Thus, the 
spatial resolution of the restored image is close to the telescope's diffraction 
limit. In addition, the one dimensional power spectrum was essentially the same 
for speckle-restored images based on $n_\mathrm{sel} = 100$ and 1600 input 
images. The G-band image depicted in the right panel of Figure~\ref{FIG08} 
serves as an example of the imaging capabilities of the \textit{GREGOR} solar 
telescope in the blue wavelength regime under very good but not excellent seeing 
conditions. Restored images of higher quality were observed with 
\textit{GREGOR}'s high-resolution imagers but with lower image acquisition rate 
so that they could not be used for this study. Many examples of high-resolution 
images were presented in \cite{Schlichenmaier2016}.


\subsection{Field Dependency of Image Quality Metrics}\label{SEC36}

\begin{figure}
\centering
\includegraphics[width=\textwidth]{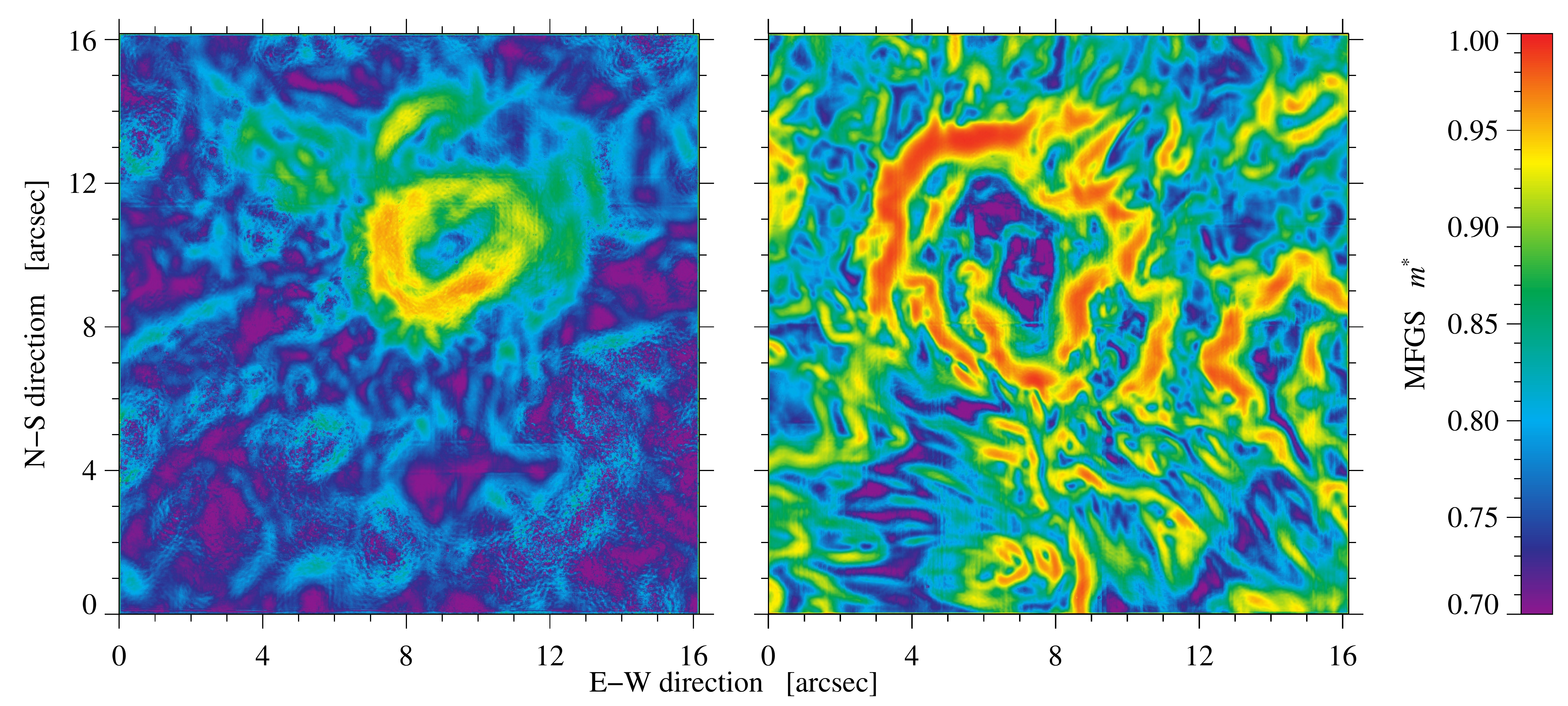}
\caption{Two-dimensional maps of the average local MFGS values $m^\ast$ for
     G-band image sequences No.~5 on 2017 June~4 (\textit{left}) and June~5
     (\textit{right}), respectively.}
\label{FIG09}
\end{figure}

As mentioned in Section~\ref{SEC31}, changing the order of summation and 
computing the similarity measure first yields an average MFGS value $m^\ast$ on 
a pixel-by-pixel basis. The corresponding average ($n_\mathrm{seq} = 40\,514$ 
and 53\,015) two-dimensional MFGS maps are depicted in Figure~\ref{FIG09} for 
G-band image sequences No.~5 observed on both June~4 and~5. Even though 
individual maps computed in this fashion are very noisy, the average maps 
strongly resemble the gradient magnitude of the average G-band images. Using the 
Scharr operator to calculate the gradient magnitude yields linear correlation 
coefficients $\rho = 0.67$ and 0.90 for the two MFGS maps in Figure~\ref{FIG09}. 
The correlation is higher, when the seeing is better. These findings reveal the 
inherent influence of the gradient magnitude operator, which is applied to both 
the original image and the median-filtered image. The superior performance of 
MFGS as an image quality metric, in contrast to simple gradient magnitude 
operators, results from the similarity measure casting the MFGS values into the 
interval $(0,\ 1]$. However, the MFGS maps in Figure~\ref{FIG09} also clearly 
demonstrate the field dependency of the MFGS image quality metric, which has to 
be considered when comparing images of different scenes on the Sun.

\begin{figure}
\centering
\includegraphics[width=\textwidth]{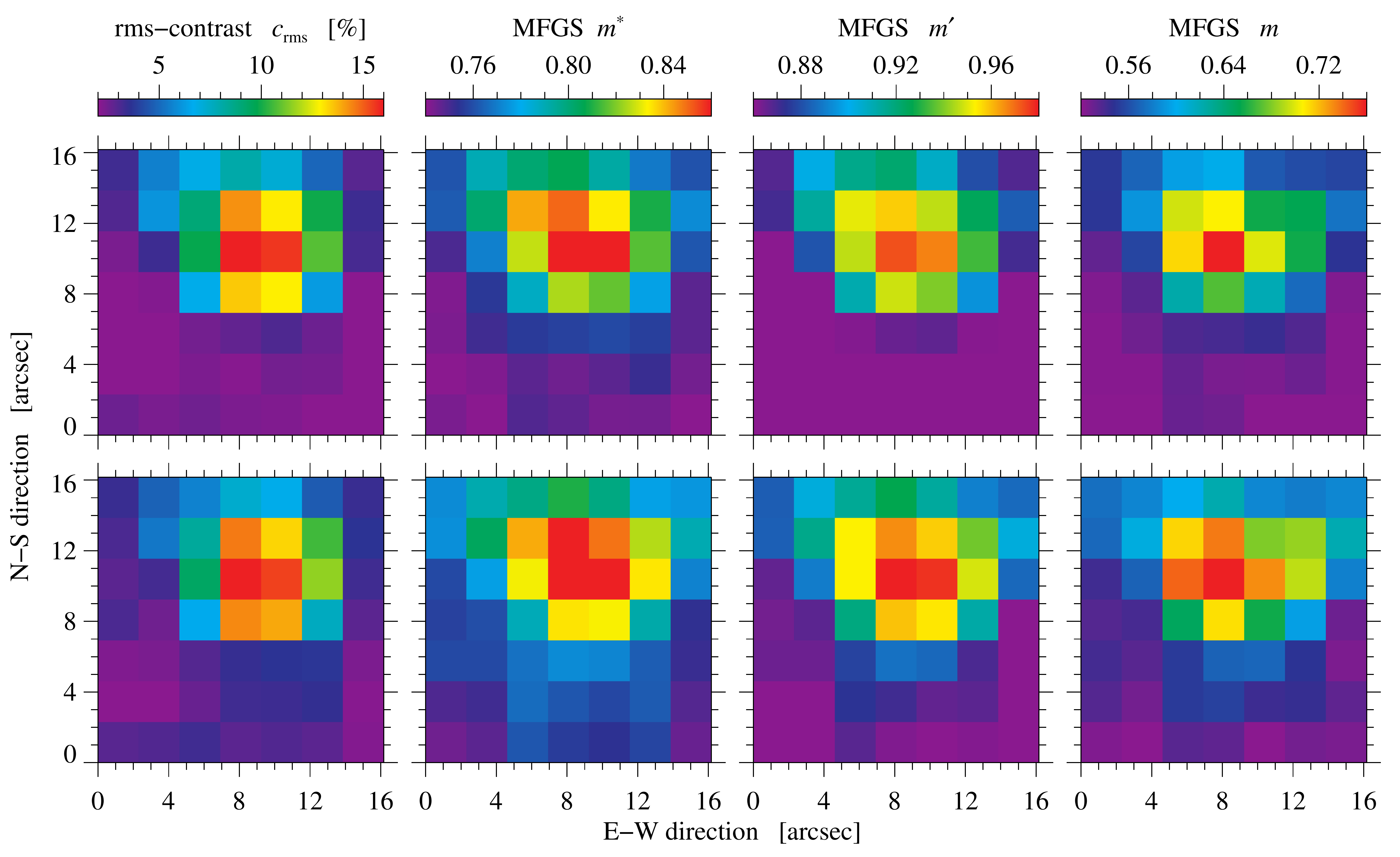}
\caption{Field dependency of the average image quality respectively seeing 
    parameters $c_\mathrm{rms}$, $m^\ast$, $m^\prime$, and $m$ 
    (\textit{left-to-right}) for the blue continuum (\textit{top}) and G-band 
    (\textit{bottom}) images computed for image sequence No.~5 on 2017 June~4. 
    The tiles have a size of $4\arcsec \times 4\arcsec$ and overlap by half. 
    Thus, there is a positional mismatch if compared to G-band images.}
\label{FIG10}
\end{figure}

Another way of assessing the field dependency is to compute the image quality 
parameters $c_\mathrm{rms}$, $m^\ast$,  $m^\prime$, and $m$ for sub-fields with 
the size of approximately the isoplanatic patch ($160 \times 160$ pixels or 
$4\arcsec \times 4\arcsec$). The results for image sequence No.~5 on June~4 are 
compiled in Figure~\ref{FIG10} for $7 \times 7$ partially overlapping 
sub-fields. Higher values occur at the location of the central pores, which was 
also used for locking the AO systems. Systematic offsets between the upper and 
lower parts of the maps are caused by the presence of several small-scale pores 
in the upper part of the images. The low-resolution MFGS maps also agree with 
the high-resolution MFGS map in the left panel of Figure~\ref{FIG09}. In 
general, all maps exhibit a close resemblance and mainly differ in the absolute 
values. Unfortunately, these data are insufficient to determine the precise 
contributions to the field dependency either due to AO correction or due to the 
presence of contrast-rich structures. The June~5 data (not shown), with a more 
dispersed appearance caused by the dominant penumbral filaments, argue for a 
stronger contribution of the contrast-rich features. To isolate the AO 
contribution, locking on granulation will likely answer this question. However, 
such \textit{GREGOR} data were not available. In many respects, the current 
results follow the trend already observed for other seeing parameters, 
\textit{e.g.} the Fried-parameter $r_0$ and the differential image motion 
\citep[\textit{e.g.}][]{Denker2005b, Denker2007a, Berkefeld2010}.

%
%

\section{Discussion}\label{SEC4} 

The current ($2560 \times 2160$ pixels) and next (4k $\times$ 4k pixels) 
generation of sCMOS cameras offer an image acquisition rate of 50~Hz in 
full-frame, global-shutter mode. This frame rate marks the point, where the MFGS 
curve $q(f_\mathrm{acq})$ becomes flatter and eventually reaches a constant 
value (Figure~\ref{FIG06}), when $f_\mathrm{acq}$ approaches the coherence time 
of the prevailing daytime seeing conditions. Considering the diminishing 
marginal benefits of $q(f_\mathrm{acq}$) for image restoration on the order of 
30\%, faster camera systems are desirable but come at a significant cost for 
data storage and processing. A possible  solution are cameras with on-chip image 
buffers and fast image-caching so that external triggering becomes possible, 
either with ultra-fast, small-format  cameras or with signals directly from the 
AO system. Ultimately, the latency between exposure time (typically 1\,--\,2~ms) 
and image acquisition time (now typically about 5\,--\,20~ms) has to be 
minimized as much as possible. 

This latency can also be interpreted as the duty cycle of photon capture, which 
is only 16\% for a image acquisition rate of 160~Hz and an exposure time of 
1~ms. In principle, the exposure time can still be somewhat increased without 
affecting the image acquisition rate. However, the shorter exposure times 
ensured that they fell below the coherence time of the seeing. The above 
mentioned image acquisition rate of 50~Hz has also implications for imaging 
spectroscopy and spectropolarimetry. Taking the \textit{GREGOR Fabry-P\'erot 
Interferometer} \citep[GFPI,][]{Denker2010a, Puschmann2012} as an example, 
typical image acquisition rates and exposure times are 10\,--15~Hz and 
10\,--\,20~ms, respectively, with a duty cycle of only 15\,--\,20\%. In this 
case, using cameras with a low read-out noise, a small full-well capacity (a few 
10\,000~e$^-$), a duty cycle close to 100\%, and an image acquisition rate of at 
least 50~Hz will lead to significant performance gains -- in particular, 
considering the ``photon-starved'' observations because of very narrow 
bandpasses (down to 2.5~pm).

Currently, both HiFI sCMOS cameras are attached to one control computer, where 
the images are saved to a RAID-0 array of eight SSDs at about 660~MB~s$^{-1}$. 
Data are already written while new images are captured in a ring buffer. In the 
standard HiFI observing mode, two times $n_\mathrm{set} = 500$ images are 
acquired in $\Delta t_\mathrm{set} = 10$~s and recorded to disk in less than  
$\Delta t_\mathrm{rec} = 20$~s, which includes overhead for real-time display of 
the blue continuum and G-band images and deferred writing from the ring buffer 
of the last images in the set. Already on site, all images undergo basic 
calibration and the best $n_\mathrm{sel} = 100$ images are selected and saved 
for \textit{post-facto} image restoration. Thus, the results of this study are 
directly applicable to standard HiFI observations. However, the cadence of the 
restored images is two times slower than in principle possible with a data 
acquisition rate of $f_\mathrm{acq} = 50$~Hz. However, in a master-slave 
configuration, where one sCMOS camera is attached to a dedicated computer, 
adequate data rates for several cameras can be achieved, while the  
synchronization is provided by the PTU of the master computer.

The image restoration was carried out on a dedicated server at AIP containing 
four CPU sockets with 16 cores each, utilizing AMD Opteron 6378 processors with 
a clock speed of 2.4~GHz. Restoring a single isoplanatic patch takes about 90~s 
on a single core using 100 input images. Some overhead for reading and preparing 
the input images for parallel processing takes about the same amount of time. 
Adding more input images to the restoration process only raises the computing 
time linearly for the isoplanatic patches while the overhead essentially remains 
the same. Thus, frame selection also minimizes the computational resources. If 
sufficient cores are available, as for example on other, larger AIP computation 
clusters, processing in almost real-time \citep{Denker2001a} becomes possible 
even for full-format 5-megapixel image sets. Pushing high-resolution imaging to 
the limit leads to instrument designs such as the \textit{Visible Broadband 
Imager} \citep[VBI,][]{McBride2012} for the 4-meter aperture \textit{Daniel K.\ 
Inouye Solar Telescope} \citep[DKIST,][]{Tritschler2016}. Details of the VBI 
image acquisition and processing pipeline are laid put in \citet{Beard2014} 
incorporating AO-corrected and frame-selected images for real-time speckle 
restoration using on-site GPU technology.

Frame selection necessarily leads to unevenly sampled restored image sequences. 
The standard deviation of the effective observing time is about 2.4~s for a 
10-second time interval, which indicates significant clustering of the moments 
with the best seeing conditions. However, on average the cadence of the restored 
image sequences remains unaffected. In most cases, it is sufficient to have 
context images with moderate cadence to follow major morphological changes or as 
input for optical flow techniques. In local correlation tracking 
\citep[LCT,][]{November1988, Verma2011}, for example, a cadence of 60~s is 
sufficient, and an unevenly sampled time-series will have no impact when 
time-averaged flow maps are computed for studying persistent flow patterns -- in 
particular, the effective observing time is known and thus the time interval 
that is needed to compute flow velocities for individual flow maps. For 
transient events and special purpose studies, such as the one at hand, all raw 
data, at the highest possible acquisition rate, can of course be kept.

The first high-cadence imaging systems for frame selection and image restoration 
became operational almost three decades ago \citep{Scharmer1989}. Video 
technology and fast frame grabbers already provided frame rates of 25\,--\,60~Hz 
at that time. Even though the dynamic range increased from 8-bit video images to 
16-bit for current sCMOS detectors, noise still remains an issue. However, the 
fixed-pattern noise of sCMOS sensors rather than photon or read-out noise has to 
be treated carefully in subsequent image restoration. Nonetheless, the statement 
of \citet{Scharmer1989} remains true that accurate photometry is not the goal of 
high-cadence imaging systems but nearly diffraction-limited imaging.

\citet{Deng2015} introduced MFGS to solar imaging and presented several short 
(200 frames) image sequences obtained in the TiO-band at $\lambda$705.8~nm. The 
images were acquired at the 1-meter aperture \textit{New Vacuum Solar Telescope} 
\citep{Liu2014} with a CMOS camera ($2560 \times 2160$ pixels, image scale 
0.04\arcsec\ pixel$^{-1}$) at a rate of 10\,--15~Hz and millisecond exposure 
times. The present study extends the original work by significantly increasing 
the statistics, introducing AO-corrected images, presenting strictly synchronous 
blue continuum and G-band images, raising the image acquisition rate by a factor 
of ten, and proposing a different, local implementation of the MFGS image 
quality metric. The results do not support the conjecture by \citet{Deng2015} 
that the simple difference operator $G_x = [-1,\ 1]$ performs better than other 
implementations of gradient operators. There is a tendency that Sobel-type edge 
enhancement operators deliver higher MFGS values. However, scatter plots show a 
very tight correlation between different MFGS implementations, demonstrating a 
strictly  monotonic though not necessarily a linear relationship. Directly 
comparing MFGS values from the original and current study proves to be 
difficult, considering different instrumental setups -- in particular, the 
seeing conditions in the blue spectral region are more challenging compared to 
the near-infrared TiO-band. An alternative approach to evaluate image quality 
metrics is numerical modelling. \citet{Popowicz2017} used high-spatial 
resolution solar images from space taken with the Japanese \textit{Hinode} 
mission \citep{Kosugi2007} and degraded them according to models of atmospheric 
turbulence. Thus, a large number of image quality metrics were quantitatively 
evaluated, with MFGS among the best performing methods.

The challenges, when high-cadence imaging meets large-format detectors, were 
already laid out for high-resolution imaging and imaging spectropolarimetry in 
\citet{Denker2010b}. Based on first-hand experience with the \textit{GREGOR} 
solar telescope, objective and robust image quality metrics are of utmost 
importance for navigating the large data volume from instruments such as GFPI 
and HiFI. In addition, the quality record provided by MFGS will not only 
facilitate searching databases, but at the same time, it will provide an 
extensive database of seeing conditions. \citet{Denker2018a} discussed data 
management and a collaborative research environment for the medium-sized 
\textit{GREGOR} project. However, major research infrastructures such as DKIST 
and the future \textit{European Solar Telescope} \citep[EST,][]{Collados2010b} 
require different approaches. The petascale cyberinfrastructure for the DKIST 
Data Center, for example, is summarized in \citet{Berukoff2016}. Yet, image 
quality metrics like MFGS continue to provide valuable metadata for the expected 
data products.

%
%

\section{Conclusions}\label{SEC5} 

The current study presents high-cadence and high-resolution observations of a 
representative sample of photospheric features including granulation, a compact 
pore, and a small sunspot with umbral dots and complex penumbral fine structure. 
The new sCMOS imaging system HiFI with two synchronized cameras allows us to 
record images in two different wavelengths simultaneously, which enabled us to 
obtain the wavelength and field dependency of image quality metrics and to 
assess the prevailing seeing conditions. An objective assessment becomes 
increasingly important, because the huge volume of imaging data generated by 
frame selection and lucky imaging necessitates an automatic inspection of image 
quality for image restoration and in database applications.

Pushing the limits of HiFI resulted in very high-cadence solar imaging at 
$f_\mathrm{acq} = 160$~Hz with millisecond exposure times. Even in these image 
series, seeing is still variable on millisecond time-scales and shows variations 
from milliseconds to about one hour -- the length of the observed time-series. 
Already in the standard full-frame mode, the data acquisition rate 
$f_\mathrm{acq} = 50$~Hz of the sCMOS detectors outperforms the existing 
PCO.4000 facility cameras at \textit{GREGOR} by a factor of ten. This is the 
minimum data acquisition rate required to reap the benefits of frame selection, 
as demonstrated when quantifying the improvement in image quality 
$q(f_\mathrm{acq})$ based on MFGS metrics in Section~\ref{SEC34}. Originally 
introduced by \citet{Deng2015}, this method was applied in this study to 
AO-corrected blue continuum and G-band images, whereby establishing that frame 
selection significantly improves the temporal coverage of high-resolution 
observations, when the seeing becomes good enough for image restoration -- in 
particular, when the seeing conditions are good to very good. The threshold $m = 
0.65$ for G-band HiFI images is a good criterion for effective speckle masking 
image restoration.

The correlation analysis in Section~\ref{SEC33} demonstrates that the three 
implementations of MFGS $m$, $m^\prime$, and $m^\ast$ behave very similar. The 
time-series of MFGS values are tightly correlated for G-band and blue continuum 
images. MFGS values for G-band images are systematically higher than for blue 
continuum images, which cannot be explained by the wavelength dependency of MFGS 
values, but is a consequence of the unique radiative transfer characteristics in 
this molecular band, which is also exploited in proxy-magnetometry of 
small-scale magnetic features. The original MFGS algorithm of \citet{Deng2015} 
has the shortest computation time but does not perform well, when the FOV 
contains features with a strong directional preference, whereas the other two 
MFGS implementations are very robust and deliver superior results for 
post-processing, when computational efficiency is not a major concern. The field 
dependency of MFGS was substantiated in Section~\ref{SEC36}, which complicates 
comparing time-series with different structural contents. However, within a 
time-series focusing on the same scene on the Sun, MFGS is a powerful tool to 
assess image quality and to identify the moments with the best seeing 
conditions. In comparison to all MFGS metrics, using the (granular) rms-contrast 
as a discriminator for image quality or seeing conditions yields unsatisfactory 
results.

In summary, the present work evaluated image quality metrics such as MFGS and 
image contrast for AO-corrected images. The present results support the notion 
that MFGS is an objective but likely not a universal image quality metric. For 
this reason, additional investigations are necessary studying scale dependency, 
center-to-limb variation, and susceptibility to noise and straylight, among 
others.

%
%


\bigskip
\noindent
\begin{minipage}{\textwidth}
\footnotesize\textbf{Acknowledgments}$\quad$
The 1.5-meter \textit{GREGOR} solar telescope was built by a German consortium 
under the leadership of the Kiepenheuer Institute for Solar Physics in Freiburg 
with the Leibniz Institute for Astrophysics Potsdam, the Institute for 
Astrophysics G\"ottingen, and the Max Planck Institute for Solar System Research 
in G\"ottingen as partners, and with contributions by the Instituto de 
Astrof\'{\i}sica de Canarias and the Astronomical Institute of the Academy of 
Sciences of the Czech Republic. We thank Drs.\ Peter G\"om\"ory and Thomas 
Granzer for carefully reading the manuscript and providing valuable comments. 
CD, CK, HB, and MV were supported by grant DE 787/5-1 of the Deutsche 
Forschungsgemeinschaft (DFG). SJGM acknowledges support of project VEGA 
2/0004/16 and is grateful for financial support from the Leibniz Graduate School 
for Quantitative Spectroscopy in Astrophysics, a joint project of the Leibniz 
Institute for Astrophysics Potsdam and the Institute of Physics and Astronomy of 
the University of Potsdam. This study is supported by the European Commission's 
FP7 Capacities Program under the Grant Agreement number 312495. \medskip

\textbf{Disclosure of Potential Conflicts of Interest}$\quad$ The authors 
declare that they have no conflicts of interest.
\end{minipage}
\newpage


%
%


\end{article}

\end{document}